\documentclass[prd,preprint,tightenlines,floatfix,showpacs,preprintnumbers,nofootinbib,eqsecnum]{revtex4}
 \usepackage[dvips,final]{graphicx}
  \usepackage{amssymb}
   \usepackage{amsmath}
    \usepackage{epsfig}
     \usepackage{bm}% bold math

\newcommand{\al}{\alpha}
\newcommand{\be}{\beta}

\newcommand{\cL}{{\cal L}}

\newcommand{\dd}{\partial}
\newcommand{\de}{\delta}
\newcommand{\di}{\displaystyle}

\newcommand{\hg}{\hat{g}}

\newcommand{\hR}{\hat{R}}
\newcommand{\hT}{\hat{T}}
\newcommand{\hvp}{\hat{\vp}}
\newcommand{\la}{\lambda}

\newcommand{\si}{\sigma}

\newcommand{\ve}{\varepsilon}
\newcommand{\vk}{{\bf k}}
\newcommand{\vp}{\varphi}

\begin{document}
\thispagestyle{empty} \preprint{\hbox{}} \vspace*{-10mm}

\title{Self-consistent invariant dynamics of scalar perturbations in the inflationary cosmology}

\author{Roman~S.~Pasechnik}
\email[E-mail:]{rpasech@theor.jinr.ru} \affiliation{Bogoliubov
Laboratory of Theoretical Physics, JINR, Dubna 141980, Russia}
\affiliation{Faculty of Physics, Moscow State University, Moscow
119992, Russia}

\author{Gregory~M.~Vereshkov}
\email[E-mail:]{gveresh@ip.rsu.ru}
\affiliation{Research Institute
of Physics, Rostov State University, Rostov-on-Don 344090, Russia}

\date{\today}

\begin{abstract}
The gauge-independent invariant approach to investigation of the
linear scalar perturbations of inflaton and gravitational fields
is developed in self-consistent way. This approach allows to
compare various gauges used by other researchers and to find
unambiguous selection criteria of physical and coordinate effects.
We have shown that the so-called longitudinal gauge commonly used
for studying the gravitational instability leads to overestimation
of physical effects due to the presence of nonphysical proper time
perturbations. Equation of invariant dynamics (EID) is derived.
The general long-wave solution of EID for an arbitrary potential
$U(\phi)$ has been obtained. We have also found analytical
solutions for all wave lengths at all stages of the universe
evolution in the framework of simplest potential
$U(\vp)=m^{2}\vp^{2}/2$. We have constructed the analytical
expressions for the energy density perturbations spectrum
$\Delta(k,t)$ at all possible $k$ and $t$. Amplitude of the
long-wave spectrum in the case of the transition from short waves
to long ones occurs at the inflationary stage is almost flat, i.e.
has the Harrison-Zeldovich form, for arbitrary potential $U(\vp).$
\end{abstract}

%Keywords:
\maketitle

\section{Introduction}

The standard inflationary model \cite{Guth,Linde0,Linde1} is the
most successful theoretical model for explanation of the
observable Universe structure. According to this model at the
early stages the Universe was in an unstable vacuum-like state
characterized by a slow linear drop of the Hubble parameter $H(t)$
with time growth thus the cosmological expansion had a
quasiexponential character
\[
  a(t)\sim \exp\int \limits_0^t H(t) dt.
\]
This stage of evolution of the Universe is called the inflationary
epoch. For reviews on inflationary cosmology see
Refs.~\cite{Linde2,Albrecht,Lazarides}.

The simplest model of the cosmological inflation for the flat
universe is described by Lagrangian
\begin{eqnarray}
  \cL=-\frac1{2\varkappa}\hR(\hg_{ik})
        &+&\frac12\hg^{ik}\hvp_{,\;k}\hvp_{, \;i}-U(\hvp).
 \label{lagran}
\end{eqnarray}
Einstein equations for this model are
\begin{eqnarray}
  \hR_i^k-\frac12\de_i^k \hR=\varkappa \hT_i^k, \qquad
  \hT_i^k&=&\hvp_{;i}\hvp^{;k}-\de_i^k\left(
  \frac12\hvp_{;l}\hvp^{;l}-U(\hvp)\right),
 \label{enstein}
\end{eqnarray}
where $\varkappa=1/M_{Pl}^2$ is the gravitational constant,
$M_{Pl}$ is the Plank mass. The equation of motion of the inflaton
field is
\begin{eqnarray}
  \hvp_{;i}^{;i}+\frac{\dd U(\hvp)}{\dd \hvp}&=&0.
 \label{inf}
\end{eqnarray}
The symbol ``\^{}'' here implies that metric $\hat{g}_{ik}$ and
inflaton (scalar) field $\hat{\vp}$ are quantum operators. We will
divide their into the classical spatially homogeneous parts
denoted $g_{ik}$ and $\vp(t)$ correspondingly and the quantum
fluctuation operators $\chi(t, \vec x)$ and $h_{ik}(t, \vec x)$,
describing small perturbations.

The main properties of the inflationary universe may be seen in
the model with quadratic potential \cite{Linde0}
\begin{eqnarray}
  U(\hvp)=\frac12m^{2}\hvp^{2}.
 \label{chaot}
\end{eqnarray}
This model is used in our work for describing
of some new aspects of the self-consistent invariant dynamics of
inflaton and gravitational fields.

Traditionally there are three main problems in the framework of
the model (\ref{lagran}). The first problem is concerned with the
self-consistent dynamics of spatially homogeneous fields
\[
  \vp(t)=\langle \hvp \rangle, \qquad g_{ik}=\langle \hg_{ik}
  \rangle= diag (1,\; -a^2(t),\;-a^2(t),\;-a^2(t)).
\]
that is described by equations (\ref{enstein}) and (\ref{inf}) in
the zeroth order of perturbation theory:
\begin{eqnarray}
\begin{array}{l} \di
  3H^2=\varkappa\left(\frac12\dot{\vp}^2
                      +U(\vp)\right),
\\[5mm]\di
  \ddot{\vp}+3H\dot{\vp}+\frac{\dd U(\vp)}{\dd \vp}=0.
\end{array}
 \label{model}
\end{eqnarray}
Here $H=\dot{a}/a$. In further calculations we'll use both the
cosmic time $t$ ($g_{00}=1$) and the conformal time $\eta$
($g_{00}=a^2$). Recall, the dot `` $\dot{ }$ '' denotes a
derivative with respect to the cosmic time, the prime `` ${}'$ ''
denotes one with respect to the conformal time.

Solution of this problem is well-known and underlies the chaotic
inflation scenario \cite{Linde0}. We have unified the
corresponding calculations. The system of background equations
(\ref{model}) is solved analytically at all times $t$.
Consideration of dynamical properties for the simplest inflation
model is necessary for solution of the second problem:

The second problem is connected with the self-consistent dynamics
of spatially inhomogeneous quantum fluctuations of inflaton and
gravitational fields
\[
  \chi(t, \vec x)= \hvp(t, \vec x) -\vp(t), \qquad h_{ik}(t, \vec
  x)=\hg_{ik}(t, \vec x)-\langle \hg_{ik} \rangle
\]
both at inflationary and postinflationary stages of the Universe
evolution. The theory of cosmological perturbations is based on
expanding the Einstein equations to linear order about the
background metric. The theory was initially developed in
pioneering works by Lifshitz \cite{Lifshitz}. Two aspects of the
problem are: (i) the stability of the inflationary process; (ii)
the back reaction of the long-length perturbations on the
expansion of the Universe in the past and in the present epoch
\cite{Abramo}.

The third problem is how the Harrison-Zeldovich spectrum
\cite{HarZel} was formed from the inflaton field vacuum
fluctuations, the length of which is much less than the Universe
size at the beginning of the inflation.  As it is known, the
solution of this problem is the basis for the modern theory of the
large scale structure formation in the Universe.

A large amount of papers is devoted to solving the last two
problems mentioned above which has been investigated by various
researches for more than twenty years. The conventional approach
is based on the investigation of the scalar perturbations in the
so-called diagonal or longitudinal gauge as the most comfortable.
The theory is formulated in terms of the relativistic scalar
potential $\Phi(\vec x,t)=\de g_{00}(\vec x,t)$; all other
observable values are expressed via $\Phi(\vec x,t)$. The subject
of the discussion, which constantly appears in the literature, is
if the predictions of the physical consequences of the theory are
invariant. Two point of view clashed at this point. Some people
suppose, that the longitudinal gauge is physically preferred
because its main object, $\Phi(\vec x,t)$, is invariant itself
\cite{Mukh}. The opposite claim is that $\Phi(\vec x,t)$ describes
effects in the fixed frame of reference \cite{Unruh,Grisch}; in
other frames of reference the observed phenomena may look in
another way.

In fact, the question is if the theory in the longitudinal gauge
can be used to reconstruct the past of the Universe on the basis
of the observations that will be performed. The Fourier-image of
$\Phi_{\vk}(t)$ is the functional  $\Phi_{\vk}\{H(t)\}$, and the
Hubble function $H(t)$ contains information about the inflaton
field and the inflationary process  itself.  The theoretical
reconstruction of the past is possible only in the case of
$\Phi_{\vk}\{H(t)\}$ is an invariant functional. On the other
hand, in the case of the information contained in
$\Phi_{\vk}\{H(t)\}$ essentially depends on the properties of the
frame of reference that is prescribed by the longitudinal gauge,
the theory of relativistic scalar potential can not be used to
interpret the results of the experiment.

Our claim --- that the theory in the longitudinal gauge is
noninvariant --- is made on the basis of the proposed invariant
dynamics of the scalar perturbations. The main result of our
investigation is the strict proof of the following statement. {\it
The invariant information about the dynamics of the scalar
perturbations is selected from the equations of the linear
gravitational instability  theory  by the identical mathematical
transformations without making use of any gauge condition at any
stage of the calculations.} Our theory is formulated as a closed
system of equations for the invariant $J_\vk$ of metric
perturbations, invariant function $\chi_{\vk {\rm inv}}$ of the
perturbations of the inflanton field, its derivative
$\dot{\chi}_{\vk {\rm inv}}$ and energy density perturbations
$\de\ve_{\rm inv}.$ The theory is such that the invariance of the
physical values follows from their mathematical definitions, and
the invariance of the equations does from the way to obtain them
without any gauge.

Invariant approach allows us to compare different gauges which are
used in the works of other researches and to find unambiguous
separation criteria of physical and coordinate effects. The
problem of such criteria existence was widely discussed also in
\cite{Linde1,Mukh,Unruh,Grisch}. We have shown that the
longitudinal gauge leads to the overestimation of the physical
effect as a result of the strong perturbations of the proper time
in frame of reference specified by the longitudinal gauge. The
general qualitative properties --- the power-law instability of
long-length perturbations and the formation of the
Harrison-Zeldovich spectrum --- are the same in the two
approaches, but the numerical values are different, the
perturbations in the longitudinal gauge theory being several times
greater. Invariant approach excludes nonphysical coordinate
effects and gives a key for analytical investigation of equations
in the perturbation theory that is the aim of our work.

Among the theories formulated in the fixed gauges, the synchronous
gauge theory proposed by Lifshitz \cite{Lifshitz} is an adequate
one, because it this gauge the evolution of the background and
perturbations is analyzed in one and the same time.

We have analytically investigated of the equation of invariant
dynamics (EID) for $J_\vk.$  The general long-wave solution of EID
as a functional of background solution for arbitrary potential
$U(\phi)$ is obtained. For one of the simplest model of inflaton
field (\ref{chaot}) we have found explicit solutions of EID for
all wave numbers $k$ and times $t$. An important point of our
investigation is that both the background characteristics and
characteristics of perturbations at the inflationary and
postinflationary stages in the considering model (\ref{lagran}) of
the Universe are completely defined by the parameter $m$ and the
initial values of the Hubble function $H_{0}$ and the scale factor
$\tilde{a}_0.$

\section{Background dynamics: an analytical glance at the Linde's chaotic inflation}

The system (\ref{model}) for potential (\ref{chaot}) turns to
\begin{eqnarray}
\begin{array}{l} \di
  \dot{H}=-\frac\varkappa2\dot{\vp}^2,
\\[5mm]\di
  \ddot{\vp}+3H\dot{\vp}+m^{2}\vp=0.
\end{array}
 \label{eqv}
\end{eqnarray}
In the limit $t\to 0$ solutions of this system can be represented
as power series
\begin{eqnarray}
  \vp(t)=\sum_{n=0}^{\infty}a_{n}t^{n}, \qquad
  H(t)=\sum_{n=0}^{\infty}b_{n}t^{n}.
 \label{sol1}
\end{eqnarray}
Substituting (\ref{sol1}) in the
(\ref{eqv}) we obtain the system of equations
\begin{eqnarray}
\begin{array}{l} \di
  \sum_{n=0}^{\infty}nb_{n}t^{n-1}=-\frac\varkappa2\sum_{k,n=0}^{\infty}kna_{k}a_{n}t^{k+n-2},
\\[5mm]\di
  \sum_{n=0}^{\infty}n(n-1)a_{n}t^{n-2}+3\sum_{k,n=0}^{\infty}na_{n}b_{k}t^{k+n-1}+
  m^{2}\sum_{n=0}^{\infty}a_{n}t^{n}=0.
\end{array}
 \label{razl}
\end{eqnarray}

\begin{figure}[!h]
 \begin{minipage}{0.49\textwidth}
  \epsfxsize=\textwidth\epsfbox{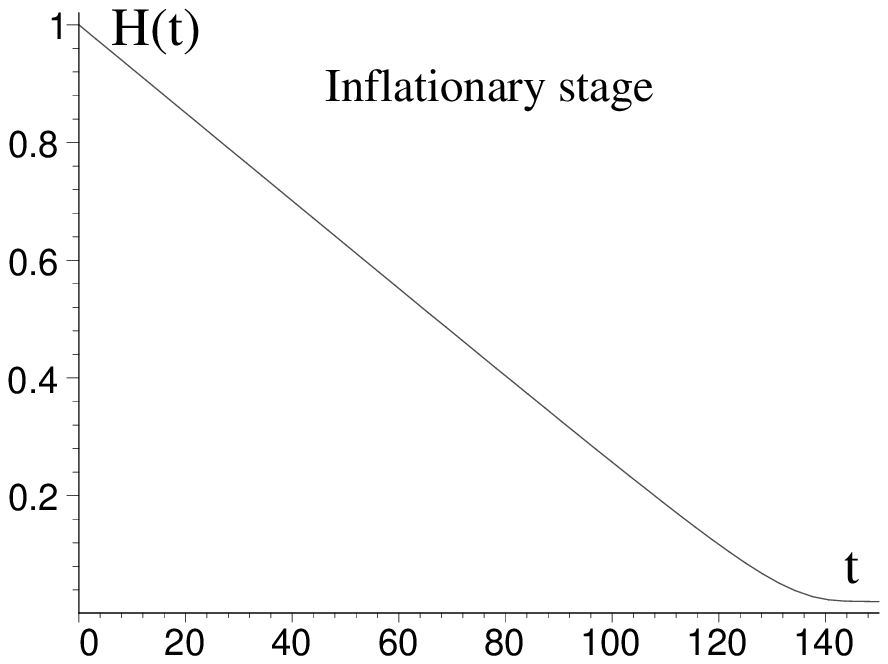}
 \end{minipage}
 \begin{minipage}{0.49\textwidth}
  \epsfxsize=\textwidth \epsfbox{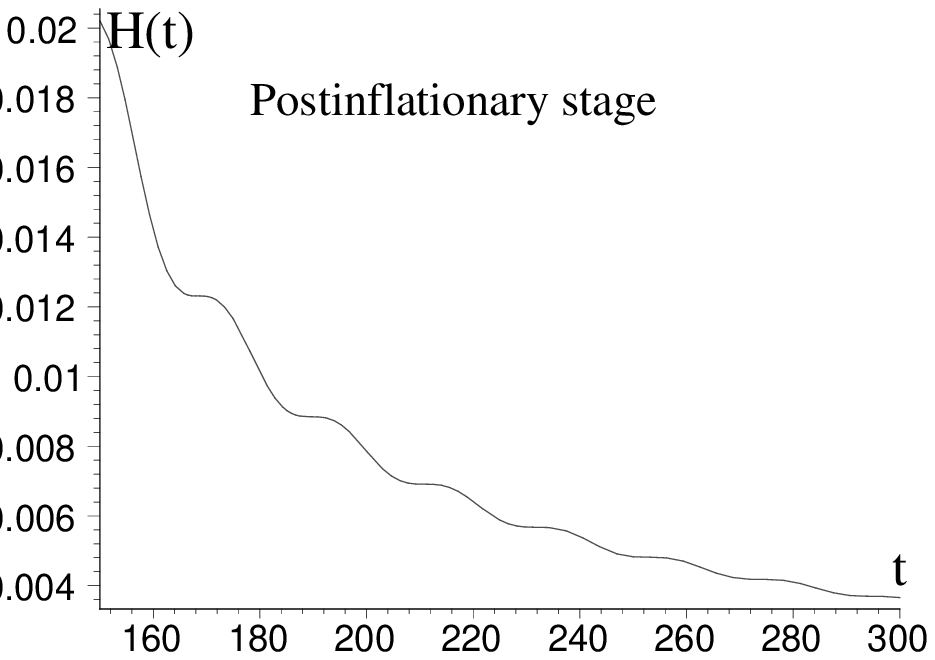}
 \end{minipage}
\caption{\small Hubble function as a function of cosmic time at
the inflationary (left) and postinflationary (right) stages;
$H_{0}=1$, $m=0.15$ (system of units $\hbar=c=8\pi G=1$),
$t_{1}\simeq 127$, $t_{0}\simeq 113$.}
 \label{fig:figH}
\end{figure}

The initial value of $H(t)$ unambiguously defines the initial
values of functions $\vp(t)$ and $\dot{\vp}(t)$. For convergence
of series we should assume that
\begin{eqnarray}
  \frac{m^2}{9\,b_{0}^2}\ll1, \qquad m\ll1, \qquad a_{2}=0.
 \label{cond}
\end{eqnarray}
Taking into account these conditions we will receive from
(\ref{razl}) the following relations for $a_{n}$ and $b_{n}$:
\[
\begin{array}{l} \di
  a_{0}=\frac{b_{0}}{m}\left(\frac6{\varkappa}\right)^{1/2}, \;
  a_{1}=-\frac{m}3\left(\frac6{\varkappa}\right)^{1/2}, \;
  a_{3}=\frac{m^{5}}{162\,b_{0}^2}\left(\frac6{\varkappa}\right)^{1/2},
  \; a_{n}=-\frac3n\,b_{0}a_{n-1} \quad \mathrm{for} \quad
  n\geqslant4;
\\[5mm]\di
  b_{1}=-\frac{\,m^{2}}3, \qquad b_{2}=0, \qquad
  b_{n}=-\varkappa{a_{1}}{a_{n}} \quad \mathrm{for} \quad
  n\geqslant3.
\end{array}
\]
Therefore the expression for expansion coefficient of function
$H(t)$ is
\[
  b_{n}=(-1)^{n+1}\:\frac{6m^{6}}{81b_{0}^2}\:\frac{(3b_{0})^{n-3}}{n!}.
\]
The series can be summed up and the result is
\begin{eqnarray}
  H(t)=b_{0}-\frac{\,m^{2}}3\:t+\frac{m^6t^2}{81b_{0}^3}-\frac{2m^6}{729b_{0}^5}
  \left(e^{-3b_{0}t}-1+3b_{0}t\right).
 \label{H}
\end{eqnarray}
For small $t$ due to conditions (\ref{cond}) this expansion
corresponds to the linear approximation -- peculiar feature of the
inflationary epoch (Fig.~\ref{fig:figH}):
\begin{eqnarray}
  H(t)=H_{0}-\frac{\,m^{2}}3\:t.
 \label{H1}
\end{eqnarray}

Let us build an analytical solution of (\ref{eqv}) in the limit
$t\rightarrow\infty,$ corresponding to the postinflationary stage.
As it is known in this limit the Hubble function $H(t)$ must
approach to $2/{3t}$ of present-day Friedman's universe. In the
equation $\ddot{\vp}+3H\dot{\vp}+m^{2}\vp=0$ one can make the
substitution $\vp=\psi/a^{3/2}$. As a result we have
\begin{eqnarray}
  \ddot{\psi}+\psi\left(m^{2}-\frac32\,\dot{H}-\frac94\,H^2\right)=0.
 \label{psii}
\end{eqnarray}
This is the equation of oscillator with the variable frequency
$\omega(t)^2=m^2+\rho(t)$, where
$\rho(t)=-\frac32\,\dot{H}-\frac94\,H^2\rightarrow0$ steadily for
$t\rightarrow\infty$. The general solution of equations like
(\ref{psii}) has the form:
\begin{eqnarray}
  \psi(t)=\frac{C_{1}}{\sqrt{2\varepsilon(t)}}\:e^{-i\int_{a}^{t}\varepsilon(t)dt}+
  \frac{C_{2}}{\sqrt{2\varepsilon(t)}}\:e^{i\int_{a}^{t}\varepsilon(t)dt},
  \quad C_{2}=C_{1}^{*}.
 \label{orr}
\end{eqnarray}
The function $\varepsilon(t)$ should be defined. Substituting
(\ref{orr}) in (\ref{psii}) and differentiating we obtain:
\begin{eqnarray}
  \frac{d}{dt}(m^2+\hat{I})D=0,
 \label{yr}
\end{eqnarray}
where $D(t)=1/\varepsilon(t)$, $\hat{I}$ is the
integro-differential operator, which is defined as
\[
  \hat{I}f=\frac14\left(\frac{d^2}{dt^2}f+2\rho{f}+
  2\int_{\infty}^{t}dt\rho\frac{d}{dt}f\right),
\]
for example, $\hat{I}\,1=\rho/2$,
$\hat{I}\rho=\frac14(\ddot{\rho}+3\rho^2).$ Equation (\ref{yr})
gives
\[
  D(t)=\frac{const}{m^2\left(1+\frac{\hat{I}}{m^2}\right)}=
  \frac1{m}\sum_{n=0}^{\infty}(-1)^{n}\left(\frac{\hat{I}}{m^2}\right)^{n}.
\]
Constant has been chosen for correspondence to zeroth
approximation of equation (\ref{psii}) -- the harmonic oscillator
with the frequency $m$. Suffice it to take into account just a
first correction. Finally we have
\begin{eqnarray}
  \varepsilon=m\left(1-\frac1{8m^2}(6\dot{H}+9H^2)\right).
 \label{eps}
\end{eqnarray}
Then instead of $\vp$ we introduce the
new function $G=\psi^2$, as a result we get
\begin{eqnarray}
  3H^2=\frac{\varkappa}{a^3}\left(\frac14\ddot{G}+m^2G-\frac34(HG)\dot{}\right).
 \label{HG}
\end{eqnarray}
On the other hand using (\ref{orr}) and (\ref{eps}) we have up to
the first order terms of $1/t$ inclusively
\begin{eqnarray}
  G\equiv\psi^2=\frac1{m}\left(1+\cos(2mt+\alpha)+\left(\frac{3H}{2m}+
  \frac9{4m}\int_{a}^{t}H^2{dt}\right)\sin(2mt+\alpha)\right).
 \label{G}
\end{eqnarray}
Substituting this expression in
(\ref{HG}) we get an equation for $H(t)$:
\[
  3H^2=\frac{\varkappa}{a^3}\left(m+\frac{3H}2\sin(2mt+\alpha)\right).
\]
This equation has been solved by substitution $a(t)=e^{\int
H(t)dt}$ then by taking the logarithm and differentiation. Up to
the second order terms of $1/t$ inclusively we get the following
asymptotic solution of the system (\ref{eqv}) for
$t\rightarrow\infty$:
\begin{eqnarray}
  H(t)=\frac2{3(t-t_{0})}+\frac1{3m(t-t_{0})^2}\sin(2m(t-t_{0})+\alpha).
 \label{Hpinf}
\end{eqnarray}
This solution corresponds to the postinflationary stage of the
Universe (Fig.~\ref{fig:figH}). Two parameters --- shift parameter
$t_{0}$ and initial phase $\alpha$ --- can be found by matching of
background solutions (\ref{H1}) and (\ref{Hpinf}).

Let $t_{1}$ is the transition time of the Universe from
inflationary to postinflationary stage. At $t=t_{1}$ we require an
execution of the following three conditions:
\begin{enumerate}
\item Continuity condition of $H(t)$:
\[
  H_{0}-\frac{\,m^{2}}3\:t_{1}=\frac2{3(t_{1}-t_{0})}+
  \frac1{3m(t_{1}-t_{0})^2}\sin(2m(t_{1}-t_{0})+\alpha);
\]
\item Smoothness condition of $H(t)$:
\[
  -\frac{\,m^{2}}3=-\frac2{3(t_{1}-t_{0})^2}+
  \frac2{3(t_{1}-t_{0})^2}\cos(2m(t_{1}-t_{0})+\alpha);
\]
\item Continuity condition for envelope of $H(t)$:
\[
  H_{0}-\frac{\,m^{2}}3\:t_{1}=\frac2{3(t_{1}-t_{0})}\,.
\]
\end{enumerate}
The solution of this system is
\begin{eqnarray}
 t_{1}=\frac{3H_{0}}{m^2}-\frac1m, \quad
 t_{0}=\frac{3H_{0}}{m^2}-\frac3m, \quad \alpha=\pi(2k+1)-4,\;
 k\in Z. \label{par}
\end{eqnarray}
In the model under consideration with $m=0.15$ in the Planck units
$\hbar=c=8\pi G=1$ the slow-rolling regime holds form $t=0$ till
$t_{1}=127$ (Fig.~\ref{fig:figH}).

For subsequent calculations we need in accurate expressions for
scale factor at all stages:
\begin{enumerate}
\item Inflationary stage $0<t\leqslant t_{1}$:
\begin{eqnarray}
\begin{array}{l} \di
 a(t)=\tilde{a}_{0}e^{H_{0}t-\frac{\beta
 t^2}{2}}=\tilde{a}_{0}e^{\frac{t(H_{0}+H(t))}{2}}, \quad
 \beta=\frac{m^2}3,
\\[5mm]\di
 a(\eta)\approx -\frac{2}{(H_{0}+H)\eta},
\end{array}
 \label{a}
\end{eqnarray}
where $\tilde{a}_0$ is the initial value of scale factor $a(t).$
We suppose that $H(\eta)\approx const$ in all differentiations and
integrations at the inflationary stage.

\item Postinflationary stage $t>t_{1}$:
\begin{eqnarray}
  a(t)\approx
  A(t-t_{0})^{2/3}, \quad a(\eta)\approx \frac{A^3}9 \eta^2, \quad
 \mathrm{where} \quad
  A=\tilde{a}_{0}e^{\frac{3H_{0}^2}{2m^2}}\left(\frac{m}2\right)^{2/3}.
 \label{apinf}
\end{eqnarray}
\end{enumerate}

Obviously the non-perturbed characteristics of the inflationary
and postinflationary stages are completely defined by parameter
$m$ and initial values of Hubble function $H_{0}$ and scale factor
$\tilde{a}_0.$

In inflationary scenario density perturbations are generated from
vacuum fluctuations of the inflaton field, for reviews see
Refs.~\cite{Mukh,Brand1,Brand2}. In the following section we
illustrate the main idea and obtain the basic formulas for the
invariants of the metric and inflaton field perturbations.

\section{Cosmological theory of scalar perturbations:
Equation of invariant dynamics.}

Proceed to the analysis of the spatially inhomogeneous quantum
fluctuations:
\[
  \chi(t, \vec x)= \hvp(t, \vec x) -\vp(t), \qquad h_{ik}(t, \vec
  x)=\hg_{ik}(t, \vec x)-\langle \hg_{ik} \rangle.
\]

Notice, the inflaton waves are the direct consequence of that the
inflaton field is described by nonlinear Klein-Gordon equation.
The inflaton itself is usually called the spatially homogeneous
mode of this field. The spatially nonhomogeneous modes will be
called below the inflaton fluctuations. In the context of scalar
field matter, the quantum theory of cosmological fluctuations was
developed by Mukhanov \cite{Mukh1}.

The physical effect under discussion is in the following.
According to the Einstein equations, the fluctuations of inflaton
field inevitably produce the potential fluctuations of
gravitational field. As a result, in the system there appears a
collective motion where the amplitudes and phases of inflaton and
gravitational wave excitations are uniquely related. The
coefficient in the relation is the rate of spatially homogeneous
inflaton change. This type of motion will be called below the
scalar inflaton-gravitational (SIG) waves. Our work is devoted to
studying SIG waves in the frame of the simplest model with the
Lagrangian (\ref{lagran}).

It is well-known, that from the scalar perturbations of metric
\cite{Mukh}
\begin{eqnarray}
 \de g_{ik}= h_{ik}= a^2(\eta) \left(
\begin{array}{cc}
 2\phi & -B_{,\al} \\
 -B_{,\al} & 2(\psi\de_{\al\be}+E_{,\al,\be})
\end{array}  \right)
 \label{hik}
\end{eqnarray}
the following invariant functions can be formed \cite{Bardeen}:
\begin{eqnarray}
\begin{array}{c} \di
  \Phi=\phi+\frac{1}{a}[(B-E')a]',
\\[3mm] \di
  \Psi=\psi-\frac{a'}{a}(B-E').
\end{array}
 \label{invari}
\end{eqnarray}

Classical problem of the linear perturbation theory ultimately
consists in search of these invariants. This problem is extremely
difficult and it hasn't been solved generally to the present day.
Many authors resort to various simplifications namely the
co-ordinates may be chosen so that the initial expression for
metric perturbations (\ref{hik}) has got more convenient form for
investigations. It is reached by using of various gauges namely
longitudinal, synchronous, co-moving gauges. But it is necessary
to understand that within any gauge there is some arbitrariness in
selecting of the co-ordinates. Therefore it is impossible to fix
co-ordinates hard within a gauge and as a result we have not a
physical effect but effects are concerned with the co-ordinates
motion itself that is the coordinates effects. Moreover in the
investigation within any gauge it is impossible to determine the
unambiguous criteria of the physical and coordinate effects
separation. Such criteria exist only in the framework of invariant
approach which is being developed in our work.

The essence of our approach involves the following. Search of two
invariants (\ref{invari}) in the general case for metric
perturbations (\ref{hik}) is the despairing problem. We have seen
that the single invariant $J$ can be constructed from the
invariants $\Phi$ and $\Psi$:
\begin{eqnarray}
  J=\Phi+\frac{1}{a}\left(\frac{a'}{a}\Psi \right)'
  =\phi+\frac{1}{a}\left(\frac{a^2}{a'}\psi \right)'.
 \label{nash_invar}
\end{eqnarray}
We will show below that for this invariant there can be obtained
and solved analytically (exactly or by using asymptotic methods)
the single second order differential equation for arbitrary
potential $U(\vp)$ without using any gauge. So, the invariant
dynamics of the SIG waves does exist.

We employ the notations, initially used by Lifshitz
\cite{Lifshitz}. The Fourier harmonics of the potential
excitations of metric are
\begin{eqnarray}
 h_i^k(\vk)=\left(
\begin{array}{cc}
 \di Q_\vk & \di  ik_{\al}\si_\vk \\
 \di  -ik^{\be}\si_\vk & \di
 \frac13(\mu_\vk+\la_\vk)\de_{\al}^{\be}
       -\frac{k_{\al}k^{\be}}{k^2}\la_\vk
\end{array}  \right).
 \label{hik_lif}
\end{eqnarray} Operations with the spatial indices
are made in all calculations with the help of 3-Euclidean metric.
One can easily see the one-to-one correspondence between the
values from (\ref{hik}) and (\ref{hik_lif}). In the longitudinal
gauge, which is used widely nowadays, $\si_\vk=\la_\vk=0$; in the
Lifshitz or synchronous gauge $Q_\vk=\si_\vk=0$. Further when we
will construct the invariant dynamics, we do not use neither of
the these gauges. However, the results of invariant dynamics will
show that the synchronous gauge has some physical advances, in
contrast to the longitudinal one.

In the first (linear) approximation the equations describe the
relation between the fluctuations of metric and the fluctuations
of the inflaton field:
\begin{eqnarray}
\begin{array}{l} \di
 \de R_i^k-\frac12\de_i^k \de R \equiv
\\[5mm] \di
 \equiv
 \frac12\left(h_i^{l;k}{}_{;l}+h^k_{l;i}{}^{;l}-h_i^{k;l}{}_{;l}
                                                 -h_{;i}^{;k}   \right)
 -h_l^kR_i^l
 -\frac12\de_i^k\left(h_{l;m}^{m;l}-h_{;l}^{;l}-h_m^lR_l^m\right)
 =\varkappa \de T_i^k\;,\:
\\[5mm] \di
 \de
 T_i^k=-h^k_m\vp_{;i}\vp^{;m}+\chi_{;i}\vp^{;k}+\chi^{;k}\vp_{;i}
 -\de_i^k\left(-\frac12h_m^l\vp_{;l}\vp^{;m}+\chi_{;l}\vp^{;l}
 -\left(\frac{\dd U}{\dd \vp}\right)_0\chi
                \right)\:.
\end{array}
 \label{enstein_fluc_tei}
\end{eqnarray} Notice, in (\ref
{enstein_fluc_tei}) there is no any restriction on the wave length
of the fluctuations.

From (\ref {enstein_fluc_tei}) we can write all equations of
linear perturbation theory:
\begin{eqnarray}
\begin{array}{c} \di
 \de R_0^0(\vk)-\de R (\vk)=\varkappa\de T_0^0(\vk),
\\[5mm] \di
 \de R_0^0(\vk)-\de R (\vk) =\frac1{a^2}\left[\frac13
 k^2N_\vk+\frac{a'}{a}M_\vk
   -3\frac{a'{}^2}{a^2}Q_\vk\right]
 =\frac{k^2}{3a^2}N_\vk+HL_\vk-3H^2Q_\vk,
\\[5mm] \di
 \de T_0^0(\vk)=\frac{\vp'}{a^2}\left(
 -\frac12Q_\vk\vp'+\chi'_\vk\right) +\left( \frac{\dd U}{\dd
 \vp}\right)_0\chi_\vk =-\frac12Q_\vk
 \dot{\vp}{}^2+\dot{\vp}\dot{\chi}_\vk +\left( \frac{\dd U}{\dd
 \vp}\right)_0\chi_\vk,
\end{array}
 \label{inv_00}
\end{eqnarray}
\begin{eqnarray}
\begin{array}{c} \di
 \de R_\al^0(\vk)=\varkappa\de T_\al^0(\vk),
\\[5mm] \di
 \de R_\al^0(\vk)=-\frac{ik_{\al}}{a^2}\left[
 \frac13N'_\vk-\frac{a'}{a}Q_\vk\right]=
 -\frac{ik_{\al}}{a}\left(\frac13\dot{N}_\vk-HQ_\vk\right),
\\[5mm] \di
 \de T_\al^0(\vk)=\frac{ik_{\al}\vp'}{a^2}\chi_\vk
   =\frac{ik_{\al}}{a}\dot{\vp}\chi_\vk,
\end{array}
 \label{inv_al0}
\end{eqnarray}
\begin{eqnarray}
 \de R_\al^\be(\vk)-\frac12 \de_\al^\be \de R (\vk)=
 \varkappa\de T_\al^\be(\vk),
 \label{inv_albe}
\end{eqnarray}
\[
\begin{array}{l} \di
 \de R_\al^\be(\vk)-\frac12 \de_\al^\be \de R (\vk)
 =\frac{1}{6a^2}\de_\al^\be\biggl[ -N''_\vk
 -2\frac{a'}{a}N'_\vk+k^2(N_\vk+3Q_\vk)+3M'_\vk+
\\[5mm] \di \hspace*{20mm}
 +6\frac{a'}{a}(M_\vk-Q'_\vk) -6\left(
 2\frac{a''}{a}-\frac{a'{}^2}{a^2}\right)Q_\vk\biggr]+
\\[5mm] \di \hspace*{20mm}
 +\frac{1}{2a^2}\frac{k_\al k^\be}{k^2}\left[
 N''_\vk+2\frac{a'}{a}N'_\vk-\frac{k^2}{3}(N_\vk+3Q_\vk)
 -M'_\vk-2\frac{a'}{a}M_\vk\right]=
\\[5mm] \di \hspace*{0mm}
 =\frac16\de_{\al}^\be\left[-\ddot{N}_\vk - 3H \dot{N}_\vk+
\frac{k^2}{a^2}(N_\vk+3Q_\vk)+3\dot{L}_\vk + 9 H L_\vk
 -6H\dot{Q}_\vk-6(2\dot{H}+3H^2)Q_\vk \right]+
\\[5mm] \di \hspace*{20mm}
 +\frac12\frac{k_\al k^\be}{k^2}\left[
 \ddot{N}_\vk+3H\dot{N}_\vk-\frac{k^2}{3a^2}(N_\vk+3Q_\vk)-\dot{L}_\vk
 -3HL_\vk\right],
\\[7mm] \di
 T_\al^\be(\vk)=-\de_\al^\be\left[ \frac{\vp'}{a^2}
 \left(-\frac12Q_\vk\vp'+\chi'_\vk \right) -\left( \frac{\dd U}{\dd
 \vp}\right)_0\chi_\vk \right]=
\\[5mm] \di \hspace*{40mm}
 =-\de_\al^\be\left[-\frac12Q_\vk\dot{\vp}^2+\dot{\vp}\dot{\chi}_\vk-
 \left( \frac{\dd U}{\dd \vp}\right)_0\chi_\vk\right].
\end{array}
\]
Here
\begin{eqnarray}
  N_\vk=\mu_\vk+\la_\vk, \quad
  M_\vk=\mu'_\vk+2ik\si_\vk, \quad
  L_\vk=\frac{M_\vk}{a}=\dot{\mu}_\vk+2\frac{ik\si_\vk}{a}.
\label{funk}
\end{eqnarray}
Equations (\ref{inv_00}) and (\ref{inv_al0}) enable to express the
values in $T_\al^\be(\vk)$ via the metric perturbations. In terms
of conformal time the value $(\dd U/\dd \vp)_0$ can be expressed
from the equation of motion for inflaton:
\[
  \frac1{\vp'}\left( \frac{\dd U}{\dd \vp}\right)_0=-\frac1{a^2}
  \left[ \frac{\vp''}{\vp'}+2\frac{a'}{a}\right].
\]
After employing this relation and constraints (\ref{inv_00}),
(\ref{inv_al0}), equation (\ref{inv_albe}) reads
\begin{eqnarray}
\begin{array}{l} \di
 \frac{1}{6}\de_\al^\be\Biggl[ -N''_\vk
 -2\frac{a'}{a}N'_\vk+k^2(N_\vk+3Q_\vk)+3M'_\vk+
\\[5mm] \di \hspace*{40mm}
 +6\frac{a'}{a}(M_\vk-Q'_\vk) -6\left(
 2\frac{a''}{a}-\frac{a'{}^2}{a^2}\right)Q_\vk+
\\[5mm] \di
 +\frac13 k^2N_\vk+\frac{a'}{a}M_\vk -3\frac{a'{}^2}{a^2}Q_\vk
 -2\left(\frac{\vp''}{\vp'}+2\frac{a'}{a}\right)\left(
 \frac13N'-\frac{a'}{a}Q_\vk\right) \Biggr]+
\\[5mm] \di
 +\frac{1}{2}\frac{k_\al k^\be}{k^2}\left[
 N''_\vk+2\frac{a'}{a}N'_\vk-\frac{k^2}{3}(N_\vk)+3Q_\vk)
 -M'_\vk-2\frac{a'}{a}M_\vk\right]=0.
\end{array}
\end{eqnarray}
The projection of these equations on the tensor basis gives two
equations for the scalar functions $N_\vk$, $M_\vk$ and $Q_\vk$:
\begin{eqnarray} N_{\vk}'{}'+2\frac{a'}{a}N_{\vk}'
   -\frac{k^2}{3}(N_{\vk}+3Q_\vk)
   -M_{\vk}'-2\frac{a'}{a}M_{\vk}=0,
\label{tri1}
\end{eqnarray}
\begin{eqnarray}
  N_{\vk}''-2\left(\frac{a'}{a}+\frac{\vp''}{\vp'} \right)N_{\vk}'
  +k^2N_{\vk}-3\frac{a'}{a}Q'_\vk
  -6\left(\frac{\vp''}{\vp'}\frac{a'}{a}-\frac{a''}{a}+\frac{a'{}^2}{a^2}
  \right)Q_\vk+3\frac{a'}{a}M_{\vk}=0.
\label{tri2}
\end{eqnarray}
Equation  (\ref{tri2}) is written in the form demonstrating the
idea of the following calculations. From Eq. (\ref{tri2}) $M_\vk$
may be expressed via $N_\vk$ and $Q_\vk$ and substituted in
(\ref{tri1}). It is the way to obtain the single equation for
$N_\vk$ and $Q_\vk$:
\[
  N_{\vk}'{}'+2\frac{a'}{a}N_{\vk}'
  -\frac{k^2}{3}(N_{\vk}+3Q_\vk)
  -M_{\vk}'\{ N_\vk,Q_k\} -2\frac{a'}{a}M_{\vk}\{ N_\vk,Q_k\}=0.
\]
It has a specific property: two functions $N_\vk$ and $Q_\vk$ in
this equation can be combined into one invariant combination only:
\begin{eqnarray}
  J_\vk=\left(\frac{a^2}{a'}N_\vk \right)'\frac1{a}-3Q_\vk,
 \label{inv_conf}
\end{eqnarray}
Coefficients of all non-invariant terms vanish via the background
equations for $a(\eta)$, $\vp(\eta)$. As a result we obtain the
single second order differential equation for the invariant
$J_\vk$ -- the equation of invariant dynamics (EID), which in
terms of the conformal time reads
\begin{eqnarray}
\begin{array}{l} \di
  J''_\vk+2\left(\frac{a''}{a'}-\frac{\vp''}{\vp'} \right)J'_\vk+
\\[4mm] \di \hspace*{10mm}
 + \left[
  k^2+3\frac{a'''}{a'}-3\frac{a''}{a}-2\frac{a''{}^2}{a'{}^2}
  +2\frac{a'{}^2}{a^2}-2\frac{\vp'''}{\vp'}+2\frac{\vp''{}^2}{\vp'{}^2}
  -2\frac{a''}{a'}\frac{\vp''}{\vp'}\right] J_\vk =0
\end{array}
 \label{ur_inv_conf}
\end{eqnarray}

Let's obtain the EID in terms of cosmic time $t$. The value $\dd
U/\dd \vp$ is suitable to express via the Hubble function by using
background system (\ref{model}):
\[
  \frac1{\dot\vp}\left( \frac{\dd U}{\dd \vp}\right)_0=
  \frac{1}{\dot\vp}\frac{\dot U}{\dot \vp}=
  -\frac12\frac{\ddot{H}}{\dot H}-3H.
\]
Using this result together with the equations (\ref{inv_00}) and
(\ref{inv_al0}), one can obtain from (\ref{inv_albe}) the
following equation:
\begin{eqnarray}
\begin{array}{l} \di
  \frac16\de_{\al}^\be\Biggl[-\ddot{N}_\vk-3H\dot{N}_\vk+
  \frac{k^2}{a^2}(N_\vk+3Q_\vk)+3\dot{L}_\vk+9HL_\vk -H\dot{Q}_\vk
  -6(2\dot{H}+3H^2)Q_\vk+
\\[5mm] \di \hspace*{10mm}
  +\frac13\frac{k^2}{a^2}N_\vk+HL_\vk+
  \left(3H^2+H\frac{\ddot{H}}{\dot{H}}\right)Q_\vk -\left(
  2H+\frac{\ddot{H}}{3\dot{H}}\right) \dot{N}_\vk \Biggr]+
\\[5mm] \di
  +\frac12\frac{k_\al k^\be}{k^2}\left[
  \ddot{N}_\vk+3H\dot{N}_\vk-\frac{k^2}{a^2}(N_\vk+3Q_\vk)-\dot{L}_\vk
  -3HL_\vk\right]=0.
\end{array}
\end{eqnarray}
The projection on tensor basis gives
\begin{eqnarray}
  \ddot{N}_\vk+3H\dot{N}_\vk-\frac{k^2}{a^2}(N_\vk+3Q_\vk)-\dot{L}_\vk
  -3HL_\vk=0,
 \label{cosm1}
\end{eqnarray}
\begin{eqnarray}
  \ddot{N}_\vk-\left(3H+\frac{\ddot{H}}{\dot H} \right)\dot{N}_\vk
  +\frac{k^2}{a}N_\vk-3H\dot{Q}_\vk -3\left(2\dot{H}-\frac{H
  \ddot{H}}{\dot{H}} \right)Q_\vk+3HL_\vk=0.
 \label{cosm2}
\end{eqnarray}
Of course, equations (\ref{cosm1}), (\ref{cosm2}) turn to
(\ref{tri1}), (\ref{tri2}) by time transformation and use of
background equations.

Now let us introduce the invariant in the following form
\begin{eqnarray}
  J_\vk=\left( \frac{N_\vk}{H} \right)^{\dot{}}
  -3Q_\vk.
 \label{inv_cosm}
\end{eqnarray}
So we find
\begin{eqnarray}
\begin{array}{c} \di
  \dot{N}_\vk=HJ_\vk+\frac{\dot H}{H}N_\vk+3HQ_\vk,
\\[4mm] \di
\ddot{N}_\vk=H\dot{J}_\vk+2 \dot{H} J_\vk
  +6\dot{H}Q_\vk+3H\dot{Q_\vk}+\frac{\ddot{H}}{H}N_\vk.
\end{array}
 \label{prois}
\end{eqnarray}

Substituting (\ref{prois}) into (\ref{cosm1}) and (\ref{cosm2}),
one obtains
\begin{eqnarray}
\begin{array}{l} \di
  H\dot{J}_\vk+(2\dot{H}+3H^2)J_\vk+(6\dot{H}+9H^2)Q_\vk
  +3H\dot{Q}_\vk+\left( 3\dot{H}+\frac{\ddot{H}}{H}\right)N_\vk-
\\[4mm] \di \hspace*{20mm}
  -\frac13\frac{k^2}{a^2}(N_\vk+3Q_\vk)-\dot{L}_\vk-3HL_\vk=0,
\end{array}
 \label{du1}
\end{eqnarray}
\begin{eqnarray}
  -3L_\vk=\dot{J}_\vk+J_\vk\left( 2\frac{\dot
  H}{H}-3H-\frac{\ddot{H}}{H}\right) -3\frac{\dot
  H}{H}N_\vk-9HQ_\vk+\frac{k^2}{a^2}\frac{N_\vk}{H}.
 \label{du2}
\end{eqnarray}
Differentiating (\ref{du2}) and substituting $\dot{L}_\vk\{J_\vk,
N_\vk, Q_\vk\}$, $L_\vk\{J_\vk, N_\vk, Q_\vk\}$ in (\ref{du1}), we
will get the equation of invariant dynamics in the cosmic time:
\begin{eqnarray}
\begin{array}{l} \di
  \ddot{J}_\vk+ \left(
  3H+2\frac{\dot{H}}{H}-\frac{\ddot{H}}{\dot{H}}\right)\dot{J}_\vk+
\\[3mm]   \di \hspace*{20mm}
  +\left(
  \frac{k^2}{a^2}+6\dot{H}+2\frac{\ddot{H}}{H}-2\frac{\dot{H}^2}{H^2}
  -\frac{\stackrel{...}{H}}{\dot{H}}
  +\frac{\ddot{H}^2}{\dot{H}^2}-3\frac{H\ddot{H}}{\dot{H}}\right)J_\vk=0.
\end{array}
 \label{ur_inv_cosm}
\end{eqnarray}
This equation may also be obtained from (\ref{ur_inv_conf}) by the
time transformation and using background equations for $\vp$ and
$a.$

Equations (\ref{ur_inv_cosm}) and (\ref{ur_inv_conf}) are valid
for any potential $U(\vp)$; a fixed potential makes function
$H(t)$ and its derivatives also fixed. The relations which connect
the invariant characteristics of the gravitational and inflaton
field perturbations are
\begin{eqnarray}
\begin{array}{c} \di
  \chi_{\vk\;{\rm inv}}=-\frac{1}{3\dot{\vp}} \left(
  HJ_\vk+\dot{H}\int\limits_0^t J_\vk dt \right),
\\[5mm]  \di
  \dot{\chi}_{\vk\;{\rm inv}}=-\frac{H}{3\dot{\vp}} \left[
  \dot{J}_\vk+\left(2\frac{\dot{H}}{H}-\frac12\frac{\ddot{H}}{\dot{H}}
  \right) J_\vk +\frac{\ddot{H}}{2H}
  \int\limits_0^t J_\vk dt \right].
\end{array}
 \label{chi_inv}
\end{eqnarray}
In particular, these equations are used to specify the initial
conditions for invariant and its derivative via those for the
perturbations of inflaton field and its derivative.

\section{Use of the gauges: longitudinal vs. synchronous}

Now let us discuss the obtained results. One can naively to assume
that the existence of the EID does not restrain us from using any
gauge, including the longitudinal one. This, however, is not
correct. Let us consider the relative perturbation of energy
density. From (\ref{inv_00}) and (\ref{inv_cosm}) one can easily
obtain
\begin{eqnarray}
  \frac{\de\ve}{\ve}=\frac{\de\ve_{inv}}{\ve}+\frac{\de\ve_{noninv}}{\ve},
  \quad\mathrm{where}
 \label{ve}
\end{eqnarray}
\[
 \begin{array}{c} \di
  \frac{\de\ve_{inv}}{\ve}=-\frac1{9H}\left[
  \dot{J}_\vk+\left(2\frac{\dot H}{H}-3H-\frac{\ddot H}{\dot
  H}\right)J_\vk \right] +\frac{\dot{H}}{3H} \int\limits_{t_0}^{t}
  J_\vk dt,
\\[5mm] \di
  \frac{\de\ve_{noninv}}{\ve}=\frac{\dot{H}}{H}\int\limits_{t_0}^{t}Q_\vk
  dt \equiv \frac{\dot{\ve}}{\ve} \de\tau,
 \end{array}
\]
where $\ve=3H^2/\varkappa$ is the background energy density,
$\de\tau=\de\int \sqrt{g_{00}} dt=const+\frac12\int Q_\vk dt$ is
the perturbation of the proper time. Analogously the noninvariant
parts of the perturbations of inflaton field and its derivative
are
\begin{eqnarray*}
\begin{array}{c} \di
\chi_{\vk\,noninv}=\dot{\phi}\de\tau_\vk(t), \qquad
\dot{\chi}_{\vk\,noninv}=
\ddot{\phi}\de\tau_\vk(t)+\frac12\dot{\phi}Q_\vk.
\end{array}
\end{eqnarray*}

Evidently the term $\de\ve_{noninv}$ does not have any physical
sense. This term reflects that if $h_{00}\ne 0$ then the time is
not synchronized in different parts of the Universe and therefore
measurements of the background energy density in different points
of the Universe lead us to the different results that is the term
$\de\ve_{noninv}$ describes not the physical change of the energy
density but the change of the clock run or time flow rate. To
escape of such nonphysical perturbations it is necessary to
analyze the background dynamics and perturbation dynamics in one
and the same time, in one and the same clock. So, it is necessary
to synchronize the clock in the whole Universe, i.e. to put
$Q_\vk=0$, i.e. to use the Lifshitz's (synchronous) gauge.

Now let us analyze physical validity of using a gauge in the
framework of invariant approach. As example we will examine the
widely used longitudinal gauge. From metric perturbations in the
general form (\ref{hik}) one can proceed to the metric
perturbations in the longitudinal gauge performing the special
coordinate transformations $\tilde{\eta}=\eta-B+E'$,
$\tilde{x}^{i}=x^{i}+E^{,i}$. Then in the new co-ordinates the
scalar metric perturbations $\tilde{\phi}=\phi+
\frac{a'}{a}(B-E')+(B-E')'$,
$\tilde{\psi}=\psi-\frac{a'}{a}(B-E')$  are invariants due to
(\ref{invari}). Metric perturbation tensor in that case is
diagonal:
\begin{eqnarray}
 h_i^k(\vk)=\left(
  \begin{array}{cc}
   \di 2\tilde{\phi} & \di  0 \\
   \di  0 & \di  -2\tilde{\psi}\de^{\al}_{\be}
  \end{array}  \right)
 \label{hik_long_full}
\end{eqnarray}
Taking into account
(\ref{funk}), the equation (\ref{cosm1}) leads to equality:
$\tilde{\psi}=\tilde{\phi}\equiv\Phi_\vk$, that is we have a
theory containing the single invariant which is named the
relativistic potential. It is presented as an advantage of the
longitudinal gauge. However, this is an illusory advantage. In the
framework of invariant approach we have established its
baselessness.

Thus, $\si_\vk =\la_\vk=0$ in the longitudinal gauge, at that
$Q_\vk\equiv 2\Phi_\vk$ and $\mu_\vk\equiv -6\Phi_\vk$. The
equation for $\Phi_\vk$ may be obtained from (\ref{cosm2}):
\begin{eqnarray}
  \ddot{\Phi}_\vk+\dot{\Phi}_\vk\left(H-\frac{\ddot{H}}{\dot{H}}\right)
  +\Phi_\vk\left(\frac{k^2}{a^2}+2\dot{H}-\frac{H\ddot{H}}{\dot{H}}\right)=0.
 \label{ur_Phi}
\end{eqnarray}
Invariant (\ref{inv_cosm}) in the
longitudinal gauge is
\begin{eqnarray}
  J_\vk=-6\left(\frac{\Phi_\vk}{H}\right)^{\dot{}}-6\Phi_\vk.
 \label{inv_long}
\end{eqnarray}
The expression for full relative energy density perturbation
(\ref{ve}) taking into account (\ref{ur_Phi}) and (\ref{inv_long})
has a form
\begin{eqnarray}
  \frac{\de\ve}{\ve}\equiv\frac{\de\ve_{inv}}{\ve}+\frac{\de\ve_{noninv}}{\ve}=
  -\frac2{3H^2}\frac{k^2}{a^2}\Phi_\vk-\frac{2\dot{\Phi}_\vk}{H}
  -2\Phi_\vk,
 \label{ve_full}
\end{eqnarray}
where the non-invariant part of perturbations is
\begin{eqnarray}
  \frac{\de\ve_{noninv}}{\ve}=2\frac{\dot{H}}{H}\int\limits_{t_0}^{t}\Phi_\vk
  dt.
 \label{non}
\end{eqnarray}
One can show that the existence of
this term in (\ref{ve_full}) is connected with an arbitrariness in
the co-ordinates selection within the gauge.

At first we consider the long-wave limit. Suppose in the
longitudinal gauge we the co-ordinates with metric
$ds^2=(1+2\Theta_\vk)dt^2-a^2(t)(1-2\Theta_\vk)\de_{ij}dx^idx^j$.
Here the metric perturbations are described by scalar function
$\Theta_\vk$. The expression for energy density perturbations is
\begin{eqnarray}
  \frac{\de\ve}{\ve}=-\frac{2\dot{\Theta}_\vk}{H}-2\Theta_\vk.
\label{Theta}
\end{eqnarray}
Now let us make some co-ordinates
transformation preserving the gauge \cite{Unruh}:
\begin{eqnarray}
  \tilde{t}=t-\la (t), \quad \tilde{x}^i=(1+\be)x^i, \quad
  \be=const.
 \label{trans}
\end{eqnarray}
Then
\[
  ds^2=(1+2\Theta_\vk+2\dot{\la})d\tilde{t}^2-
  a^2(\tilde{t})(1-2\Theta_\vk-2\be+2H\la)\de_{ij}d\tilde{x}^id\tilde{x}^j,
\]
so $\tilde{\phi}=\Theta_\vk+\dot{\la}$ and
$\tilde{\psi}=\Theta_\vk+\be-H\la$ are metric perturbations in the
new co-ordinates. Requirement of
$\tilde{\Theta}_\vk\equiv\tilde{\psi}=\tilde{\phi}$ gives the
following expression for function $\la (t)$:
\[
  \la (t)=\frac1{a}\left(\be\int a dt+\gamma\right), \quad
  \gamma=const.
\]
Therefore
\[
  \tilde{\Theta}_\vk\equiv\Theta_\vk+\de\Theta_\vk=\Theta_\vk-\frac{H}{a}\left(\be\int
  a dt+\gamma\right)+\be.
\]
Then instead of $\Theta_\vk$ we introduce a new function
$\tilde{\Theta}_\vk$ into the right hand side of expression
(\ref{Theta}). As a result we have
\begin{eqnarray}
  \frac{\de\tilde{\ve}}{\ve}=-\frac{2\dot{\Theta}_\vk}{H}-2\Theta_\vk
  +2\frac{\dot{H}}{H}\frac1{a}\left(\be\int a dt+\gamma\right).
 \label{nov_Theta}
\end{eqnarray}
Obviously the expression for
energy density perturbations in the new co-ordinates differs from
corresponding one in the old co-ordinates by the value
\[
  \Delta=2\frac{\dot{H}}{H}\int\limits_{t_0}^{t}\de\Theta_\vk dt.
\]
Therefore we see that the expression
\[
  \frac{\de\ve_{inv}}{\ve}=-\frac{2\dot{\Theta}_\vk}{H}-2\Theta_\vk-
  2\frac{\dot{H}}{H}\int\limits_{t_0}^{t}\Theta_\vk
\]
is an invariant for any coordinate transformations of the form
(\ref{trans}).

\begin{figure}[h]
 \centerline{\includegraphics[width=0.59\textwidth]{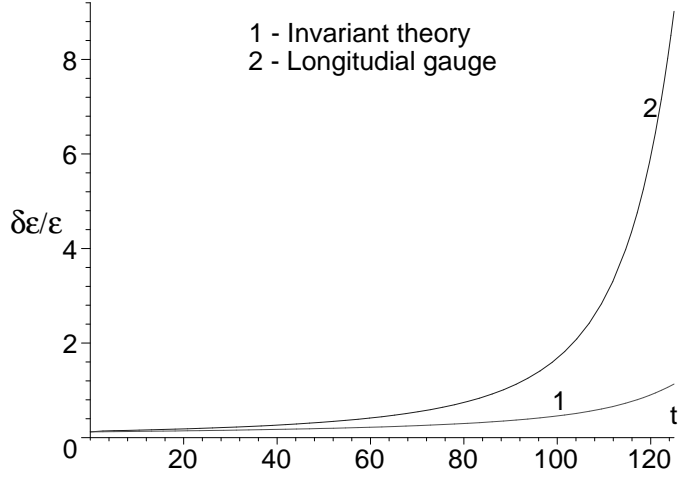}}
   \caption{\label{fig:sravn_th}
   \small Long-wave perturbations of energy density calculated at
the inflationary stage in the both approaches -- in the invariant
theory (1) and by using the longitudinal gauge (2).}
\end{figure}

Our calculations clearly prove that the use of an arbitrariness in
the co-ordinates selection for the interpretation of physical
effects is incorrect. In the longitudinal gauge together with real
physical effects there are the effects, caused by motion of the
frame of reference within the gauge. Selection and removing of
such nonphysical coordinate effects is possible only in the
framework of invariant approach. At Fig.~\ref{fig:sravn_th} one
has shown the qualitative comparison of long-wave energy density
perturbations calculated at the inflationary stage in the both
approaches. We have used the simplest model with the potential
(\ref{chaot}) in the case $m=0.15,\;H_{0}=1,\;\tilde{a}_{0}=1$.
Analytical expressions for $\de\ve_{inv}/\ve$ are shown in
Appendix.

But if in principle we can analyze the long-wave metric
perturbations in the longitudinal gauge, because we know a form of
nonphysical terms (\ref{non}), then the similar analysis of the
short-wave perturbations in this gauge is absolutely impossible.
Let's consider this problem in detail.

Invariant theory gives the single expression for nonphysical terms
(\ref{non}), which is correct for all wave lengths. Therefore in
the longitudinal gauge the expression
\begin{eqnarray}
  \frac{\de\ve_{inv}}{\ve}=-\frac2{3H^2}\frac{k^2}{a^2}\Phi_\vk-
  \frac{2\dot{\Phi}_\vk}{H}-2\Phi_\vk
  -2\frac{\dot{H}}{H}\int\limits_{t_0}^{t}\Phi_\vk dt
 \label{inv_k}
\end{eqnarray}
is an invariant thus it should be correct the following expression
\begin{eqnarray}
  -\frac2{3H^2}\frac{k^2}{a^2}\de\Phi_\vk-
  \frac{2\dot{(\de\Phi_\vk)}}{H}-2\de\Phi_\vk
  -2\frac{\dot{H}}{H}\int\limits_{t_0}^{t}\de\Phi_\vk dt\equiv0,
 \label{prot}
\end{eqnarray}
where $\de\Phi_\vk$ is the metric
change in transition to the new frame of reference with the gauge
conservation. At the same time a function $\de\Phi_\vk$ has to
satisfy the equation (\ref{ur_Phi}). We have shown that the last
condition does not hold for short waves.

Let us denote $\de\Phi_\vk=\Omega$ and consider the relation
(\ref{prot}) as equation for $\Omega$
\[
 \dot{\Omega}+\frac1{3H}\frac{k^2}{a^2}\Omega+H\Omega
 +\dot{H}\int\limits_{t_0}^{t}\Omega dt=0.
\]
We divide it now by $\dot{H},$ differentiate, and then multiply by
$\dot{H}$. As a result we have the equation
\[
  \ddot{\Omega}+\dot{\Omega}\left(H-\frac{\ddot{H}}{\dot{H}}\right)
  +\Omega\left(2\dot{H}-\frac{H\ddot{H}}{\dot{H}}\right)
  +\dot{H}\frac{d}{dt}\left(\frac1{3H\dot{H}}\frac{k^2}{a^2}\Omega\right)=0,
 \label{ur_Omega}
\]
which differs from equation for metric perturbations in the
longitudinal gauge (\ref{ur_Phi}) by the term containing
${k^2}/{a^2}$. What is the reason? Probably the longitudinal gauge
is restricted itself and an important information inevitably is
lost in transition to this gauge.

Results of SIG-waves investigations fulfilled in the various
gauges are substantially different. For example, in the
synchronous gauge we have \cite{Grisch}:
\begin{eqnarray}
 \begin{array}{c}
  \di
  ds^2=-a^2[d\eta^2-(\de_{\al\be}+h_{\al\be})dx^{\al}dx^{\be}],
 \\[5mm] \di
  g_{00}=-a^2, \; g_{0\al}=0, \; g_{\al\be}= a^2[(1+hG)\de_{\al\be}+
  h_{l}k^2G_{,\al,\be}],
  \end{array}
 \label{sinx}
\end{eqnarray}
where $G$ is the scalar function defined in the 3-space. It
satisfies equation ${G^{,\al}}_{,\al}+k^2G=0$. Since
$h^{0}_{0}\equiv Q_{\vk}=0,$ the non-invariant part of full
perturbations of energy density (\ref{ve}) is equal to zero for
any co-ordinates transformations preserving the gauge and thus the
calculation of $\de\ve/\ve$ gives a proper result in the
synchronous gauge in contrast to the longitudinal one.

The closed system of equations for metric perturbations $h$ and
$h_{l}$ is obtained by Grischuk \cite{Grisch}. The equation for
$h$ is the third order differential equation and it contains the
nonphysical solution $h\sim H$, which may be obtained by the
coordinate transformations preserving the gauge (\ref{sinx}). Note
that this solution is trivial for the same equation but written in
terms of invariant (\ref{inv_cosm}), and it is easily to show that
this equation coincides with the equation of invariant dynamics
(\ref{ur_inv_cosm}).

Now let us to say some words about the quantum perturbation theory
as it is. First of all, notice, there is a formal problem in the
theory: what function is the object of quantization. The pose of
this problem is induced by our understanding that the coordinate
effects that can be eliminated by the appropriate choice of a
classical frame of reference must not to be subject to
quantization. In the longitudinal gauge theory the discussion of
this problem is again reduced to the question whether the
description in terms of the relativistic potential is invariant or
not. One of the achievements of our invariant theory is that this
question is uniquely solved. Only the invariant metric function
which is uniquely connected with the invariant characteristics of
the inflaton field is the subject of quantization.  For the short-
wave-length metric fluctuations the normalization of quantum
operator is easy to find:
\begin{eqnarray}
  \hat{J}_\vk\approx \frac{3\dot{\vp}}{H}\hat\Psi_\vk, \quad
  \hat{\Psi}_\vk\approx \frac{1}{\sqrt{2k}a} \left[ \hat{c}_\vk
  \exp\int\limits_0^t \frac{k\, dt}{a} +\hat{c}^+_{-\vk} \exp
  \left(-\int\limits_0^t \frac{k\, dt}{a}\right)
  \right],
 \label{quant}
\end{eqnarray} where
$\hat{c}_\vk$ and $\hat{c}^+_{-\vk}$ are annihilation and creation
operators in quantum field theory. Notice, expressions
(\ref{quant}) are used only to specify the initial conditions; the
quantum dynamics itself is described by the exact operator
equation (\ref{ur_inv_cosm}). The subject of calculations is the
value of energy density fluctuations averaged over the Heisenberg
vacuum specified in the beginning of the inflation:
\begin{eqnarray}
  \left| \frac{\de\ve_{\vk\, {\rm inv}}}{\ve} \right| \equiv
  \sqrt{\bigl\langle 0| \left( \frac{\de\ve_{\vk\, {\rm inv}}}{\ve}
  \right)^2 | 0 \bigr\rangle}.
 \label{dequ}
\end{eqnarray}
In order to calculate the value (\ref{dequ}) one can solve
equation (\ref{ur_inv_cosm}) under $|c_\vk|=|c_{-\vk}^+|=1/2$. The
averaging over the phases of the complex numbers  $c_\vk$,
$c_{-\vk}$ and taking the absolute value corresponds to the
procedure of quantum averaging (see Appendix).

\section{Analytical investigation of the EID.}

At the postinflationary stage the Hubble function $H(t)$
oscillates and there are points where $\dot{H}=0$. Thus formally
the EID (\ref{ur_inv_cosm}) has to have the mathematical
singularities that could be an obstacle in investigations both
analytical and numerical. However the transformations of EID
coefficients together with background equations for arbitrary
potential (\ref{model}) have shown an absence of the mathematical
singularities in reality that has opened the possibility for
analytical investigations of the EID. This internal property of
the invariant dynamics is its significant advantage.

Let us introduce a new function $\Psi_{\vk}(t)$, related with
invariant $J_{\vk}(t)$ by relation:
\begin{eqnarray}
  J_{\vk}=\frac{3\sqrt{-\dot{H}}}{H}\Psi_{\vk}.
 \label{psi}
\end{eqnarray}
Equation for $\Psi_{\vk}(t)$ has a form:
\begin{eqnarray}
\begin{array}{l} \di
  \ddot{\Psi}_\vk+ 3H\dot{\Psi}_{\vk}+\left(
  \frac{k^2}{a^2}+\frac14\frac{\ddot{H}^2}{\dot{H}^2}-\frac12\frac{\stackrel{...}{H}}{\dot{H}}
  +2\frac{\ddot{H}}{H}-2\frac{\dot{H}^2}{H^2}-\frac32\frac{H\ddot{H}}{\dot{H}}+3\dot{H}
  \right)\Psi_\vk=0.
\end{array}
 \label{ur_inv_cosm1}
\end{eqnarray}
This equation is more convenient for analysis than initial
(\ref{ur_inv_cosm}), since all terms with singularities are
concentrated in the single coefficient by $\Psi_{\vk}$.

Now let us consider the elimination mechanism of these
singularities in detail. The background system (\ref{model}) for
arbitrary potential can be written in the form
\[
\begin{array}{l} \di
  \dot{H}=-\frac\varkappa2\dot{\vp}^2,
\\[5mm]\di
  \ddot{\vp}+3H\dot{\vp}+\frac{\dd U(\vp)}{\dd \vp}=0.
\end{array}
\]
First equation has to be differentiated with respect to $t$ twice,
the second one -- once, then obtained expression has to be
multiplied by $\varkappa\dot{\vp}$. As a result we have the
following necessary relations:
\[
\begin{array}{l} \di
  \dot{H}=-\frac\varkappa2\dot{\vp}^2, \quad
  \ddot{H}=-\varkappa\dot{\vp}\ddot{\vp},\quad
  \stackrel{...}{H}=-\varkappa\dot{\vp}\ddot{\vp}^2-\varkappa\dot{\vp}\stackrel{...}{\vp},
\\[5mm]\di
  \varkappa\dot{\vp}\stackrel{...}{\vp}+3\varkappa\dot{\vp}^2\dot{H}+
  3\varkappa\dot{\vp}\ddot{\vp}H+\varkappa\dot{\vp}^2\frac{\dd^2
  U(\vp)}{\dd \vp^2}=0.
\end{array}
\]
Expressing from first three relations the all derivatives of
function $\vp$ through derivatives of $H$ and substituting their
to the forth one, after division by $2\dot{H}$ we obtain the
following identity:
\begin{eqnarray}
  \frac14\frac{\ddot{H}^2}{\dot{H}^2}-\frac12\frac{\stackrel{...}{H}}{\dot{H}}
  -3\dot{H}-\frac32\frac{H\ddot{H}}{\dot{H}}\equiv\frac{\dd^2
  U(\vp)}{\dd \vp^2}.
 \label{osob}
\end{eqnarray}
Evidently, the structure of background
equations allows to combine the terms with singularities into the
smooth function that is the singularities cancel. Taking into
account (\ref{osob}) the EID gains the simple form:
\begin{eqnarray}
  \ddot{\Psi}_\vk+ 3H\dot{\Psi}_{\vk}+\left(
  \frac{k^2}{a^2}+\frac{\dd^2 U}{\dd
  \vp^2}+2\frac{\ddot{H}}{H}-2\frac{\dot{H}^2}{H^2}+6\dot{H}
  \right)\Psi_\vk=0.
 \label{ur_inv_cosm2}
\end{eqnarray}
One can rewrite its in the form:
\begin{eqnarray}
  \ddot{\Psi}_\vk+
  3H\dot{\Psi}_{\vk}+\left(\frac{k^2}{a^2}+U''_{\vp\vp}
  +\rho(t)\right)\Psi_\vk=0,
 \label{ur_inv_cosm3}
\end{eqnarray}
where
$U''_{\vp\vp}\equiv\frac{\dd^2 U}{\dd\vp^2},$
$\rho(t)\equiv\frac{d}{dt}\left(2\frac{\dot{H}}{H}+6H\right).$

Let us introduce the characteristic function
\[
q(k,t)=\frac{k^2}{a^2(U''_{\vp\vp}(\vp(t))+\rho(t))}.
\]
Function $q(k,t)$ has a substantially different form at various
stages of evolution but at the all applicable domain $t\in
(0\;.\,.\;\infty)$ it decreases with the time growth. There are
three types of metric perturbations depending on value $q(k,t)$:
\begin{enumerate}
\item Asymptotically long-wave perturbations, if
\begin{eqnarray} q(k,t)\ll 1; \label{long} \end{eqnarray}
\item Perturbations of middle wave lengths, if:
\begin{eqnarray} q(k,t)\sim 1; \label{mid} \end{eqnarray}
\item Asymptotically short-wave perturbations, if
\begin{eqnarray} q(k,t)\gg 1. \label{short} \end{eqnarray}
\end{enumerate}

Consider the long-wave limit (\ref{long}). The equation of
invariant dynamics (\ref{ur_inv_cosm2}) in this limit reads
\begin{eqnarray}
  \ddot{\Psi}_\vk+
  3H\dot{\Psi}_{\vk}+\left(\frac{\dd^2 U}{\dd \vp^2}
  +2\frac{\ddot{H}}{H}-2\frac{\dot{H}^2}{H^2}+6\dot{H}
  \right)\Psi_\vk=0.
 \label{ur_inv_cosm4}
\end{eqnarray}
Taking into
account (\ref{osob}) it is easily to show that
$\Psi_{\vk}(t)=\sqrt{-\dot{H}}/H$ is the particular solution of
this equation. Therefore the general solution of the EID for long
waves with the arbitrary potential $U(\vp)$ has a form:
\begin{eqnarray}
  \Psi_{\vk}(t)=C_{1}\frac{\sqrt{-\dot{H}}}{H}+
  C_{2}\frac{\sqrt{-\dot{H}}}{H}\int_{0}^{t}\frac{H^2}{\dot{H}a^3}dt.
 \label{or}
\end{eqnarray}

Deeper analysis is possible only in the framework of fixed
potential. Further calculations are fulfilled for the simplest
inflation model with the potential (\ref{chaot}). In this model
function $\Psi_{\vk}$ for long-wave metric perturbations at the
postinflationary stage oscillates with the stationary amplitude:
\begin{eqnarray}
  \Psi_{\vk}(t)\sim \sin(m(t-t_{0})+\al/2),
 \label{long_chaot}
\end{eqnarray} where $\al$ is defined in (\ref{par}). This result
may be also obtained by direct solution of the EID which in that
case reduces to the Matieu-like equation where there is a
time-dependent coefficient by the oscillating correction in the
frequency. Within that context the form of relation
(\ref{long_chaot}) means the presence of the parametric resonance
for long-wave metric perturbations at the postinflationary stage.

Let's return to initial equation (\ref{ur_inv_cosm3}) and proceed
to the conformal time $\eta$:
\begin{eqnarray}
  \Psi''_\vk+2\frac{a'}{a}\Psi'_\vk+(k^2+a^2U''_{\vp\vp}
  +a^2\rho(\eta))\Psi_\vk=0.
 \label{etur}
\end{eqnarray}
By substitution
$\Psi_\vk=\xi\Psi_{\vk}^{0}$, where $\xi$ is a new unknown
function, $\Psi_{\vk}^{0}=\sqrt{-\dot{H}}/H$ is the particular
long wave solution, this equation is reduced to
\begin{eqnarray}
  \xi''+2\xi'\frac{(a\Psi_{\vk}^{0})'}{a\Psi_{\vk}^{0}}+k^2\xi=0.
 \label{xi}
\end{eqnarray}
At the inflationary stage this equation
can be reduced to the Bessel equation for any $k$:
\[
  \xi''-2\frac{\left(1+\frac{m^2}{3H^2}\right)}{\eta}\xi'+k^2\xi=0.
\]
Its exact solution is
\[
  \xi(\eta)=\eta^{\left(1+\frac{m^2}{3H^2}+\frac12\right)}
  \left[C_1\,\mathrm{BesselY}\left(1+\frac{m^2}{3H^2}+\frac12,k\eta\right)+
  C_2\,\mathrm{BesselJ}\left(1+\frac{m^2}{3H^2}+\frac12,k\eta\right)\right].
\]
At the most part of inflationary stage the inequality
$\frac{m^2}{3H^2}\ll 1$ is valid. In this case we have
\[
  \xi(\eta)=L_{1}(k\eta \cos(k\eta)-\sin(k\eta))+L_{2}(k\eta
  \sin(k\eta)+\cos(k\eta)).
\]
Therefore the general solution of the EID at the inflationary
stage has a form
\begin{eqnarray}
\begin{array}{l} \di
  \Psi_\vk(t)=L_{1}\frac{\sqrt{-\dot{H}}}{H}\left(\int\frac{k}{a}dt\;
  \cos\left(\int\frac{k}{a}dt\right)-\sin\left(\int\frac{k}{a}dt\right)\right)+
\\[5mm]\di
  +L_{2}\frac{\sqrt{-\dot{H}}}{H}\left(\int\frac{k}{a}dt\;
  \sin\left(\int\frac{k}{a}dt\;\right)+\cos\left(\int\frac{k}{a}dt\right)\right).
\end{array}
 \label{refinf}
\end{eqnarray}
Entry conditions for $\Psi_\vk(t)$
\[
  \Psi_{\vk_0}=\sqrt{\frac{\varkappa}{2k}}\frac1{\tilde{a}_{0}}, \quad
  \dot{\Psi}_{\vk_0}=\sqrt{\frac{\varkappa k}{2}}\frac1{\tilde{a}_{0}^2}
\]
give the following expressions for integration constants $L_1$ and
$L_2$:
\begin{eqnarray}
  L_1=-H_{0}^2\sqrt{\frac{3\varkappa}2}\frac{\cos(k\eta_{0})-
  \sin(k\eta_{0})}{mk^{3/2}},\quad
  L_2=-H_{0}^2\sqrt{\frac{3\varkappa}2}\frac{\cos(k\eta_{0})+
  \sin(k\eta_{0})}{mk^{3/2}},
 \label{L12}
\end{eqnarray}
where
\[
  \eta_0\equiv\eta(t=0)=\int \frac{dt}{a} \biggr\vert_{t=0}\approx
  -\frac1{\tilde{a}_{0}H_{0}}.
\]

The equation $q(k,t)=1$ represents relation between $k$ and $t$ at
the boundary that separates the evolutional stages. Its solution
$\tilde{t}(k)$ is the time of transition from short waves to long
ones for every $k$. There are two variants depending on $k$:
\begin{enumerate}
\item The transition from short-wave perturbations to long-wave ones occurs
at the inflationary stage that is $0<\tilde{t}(k)<t_{1}$;
\item The transition from short-wave perturbations to long-wave ones occurs
at the postinflationary stage that is $\tilde{t}(k)>t_{1}$.
\end{enumerate}
For every variant the EID (\ref{ur_inv_cosm2}) at the
postinflationary stage should be solved separately.

In the first case at the times $t>t_1$ there are long-wave metric
perturbations described by (\ref{or}) or (\ref{long_chaot}). After
matching with solution (\ref{refinf}) at the moment $t_{1}$ when
the inflation terminates, we obtain
\begin{eqnarray}
  \Psi_{\vk}(t)=-H_{0}^2\sqrt{\frac{3\varkappa}2}\frac{\cos(k\eta_{0})+
  \sin(k\eta_{0})}{mk^{3/2}} \frac{\sqrt{-\dot{H}}}{H}.
 \label{refpinf}
\end{eqnarray}

Finally consider the second particular case. The short-wave
solution at the inflationary stage is (\ref{refinf}). Let us
obtain the short-wave solution at the postinflationary stage. Let
us introduce a new function $\phi=a\Psi_\vk$ into the
(\ref{etur}). Using (\ref{apinf}), we get
\begin{eqnarray}
  \phi''+\phi\left(k^2-\frac{2}{\eta^2}+\frac{m^2A^6}{81}\eta^4-\frac{4mA^3}{3}\eta
  \sin\left(\frac{2mA^3}{27}\eta^3+\be\right)\right)=0.
 \label{compl}
\end{eqnarray}
In the short-wave limit on right $t\rightarrow t_{1}$ this
equation can be simplified:
\begin{eqnarray}
  \phi''+\phi\left(k^2+\frac{m^2A^6}{81}\eta^4\right)=0.
 \label{compl1}
\end{eqnarray}
Obviously this equation is an oscillator-like equation and it
should be solved by the asymptotic method which has described in
the Section 2. One has obtained the following general asymptotic
solution:
\begin{eqnarray}
\begin{array}{l} \di
  \Psi_\vk(t)=\frac1{(t-t_{0})^{2/3}\sqrt{1+\frac{m^{4/3}}{2p^2}(t-t_{0})^{4/3}-
  \frac{3m^{8/3}}{8p^4}(t-t_{0})^{8/3}}}\times
\\[5mm]\di
  \times\biggl[M_{1}\cos\left(3pm^{1/3}(t-t_{0})^{1/3}+
  \frac{3m^{5/3}}{10p}(t-t_{0})^{5/3}-\frac{m^3}{8p^3}(t-t_{0})^3\right)+
\\[5mm]\di
  M_{2}\sin\left(3pm^{1/3}(t-t_{0})^{1/3}+
  \frac{3m^{5/3}}{10p}(t-t_{0})^{5/3}-\frac{m^3}{8p^3}(t-t_{0})^3\right)
  \biggr], \quad \mathrm{where} \; p=p(k),
\end{array}
 \label{shortresh}
\end{eqnarray}
\begin{eqnarray}
\begin{array}{l} \di
  M_{1}=-\frac{1}{m^{2/3}}\sqrt{\frac{2\varkappa}{p}}
  \frac{18H_{0}^2}{{\tilde{a}_{0}}^{3/2}m^{5/2}exp\left(\frac{9H_{0}^2}{4m^2}\right)}\left[
  \cos\left(\frac{pme^{\frac{3H_{0}^2}{2m^2}}}{2^{2/3}H_{0}}\right)+
  \sin\left(\frac{pme^{\frac{3H_{0}^2}{2m^2}}}{2^{2/3}H_{0}}\right)\right],
\\[5mm]\di
  M_{2}=-\frac{1}{m^{2/3}}\sqrt{\frac{2\varkappa}{p}}
  \frac{18H_{0}^2}{{\tilde{a}_{0}}^{3/2}m^{5/2}exp\left(\frac{9H_{0}^2}{4m^2}\right)}\left[
  \cos\left(\frac{pme^{\frac{3H_{0}^2}{2m^2}}}{2^{2/3}H_{0}}\right)-
  \sin\left(\frac{pme^{\frac{3H_{0}^2}{2m^2}}}{2^{2/3}H_{0}}\right)\right].
\end{array}
\end{eqnarray}

The equation (\ref{ur_inv_cosm2}) for function
$\chi=\Psi_\vk(t-t_{0})$ has a form:
\begin{eqnarray}
  \chi''+\left(1+\frac{p^2}{\tau^{4/3}}-\frac{4}{\tau}\sin(2\tau+\al)\right)\chi=0.
 \label{Matie1}
\end{eqnarray}
Here $\tau=m(t-t_{0})$ is the new
dimensionless variable. Depending on $\tau$ part of frequency
\[
  \rho(\tau)=\frac{p^2}{\tau^{4/3}}-\frac{4}{\tau}\sin(2\tau+\al),
  \quad \rho(\tau)\rightarrow 0 \quad \mathrm{for} \quad \tau \rightarrow
  \infty
\]
contains both monotonous and oscillating constituents thus this
equation differs from both the oscillator-like equation and the
Mathieu-like equation.

With the time growth the term ${p^2}/{\tau^{4/3}}$ drops faster
then the absolute value of $(4/\tau)\sin(2\tau+\al)$. The
condition of their equality gives an estimation of the lower time
border for the asymptotically long waves:
\begin{eqnarray}
  \frac{p^2}{\tau_{long}^{4/3}}\sim\frac1{\tau_{long}}\;
  \longrightarrow \; \tau_{long}\sim p^6.
\end{eqnarray}
From (\ref{Matie1}) one can estimate also the upper time border of
the asymptotically short waves:
\begin{eqnarray}
  \frac{p^2}{\tau_{short}^{4/3}}\sim 1\; \longrightarrow \;
  \tau_{short}\sim p^{3/2}.
 \label{csh}
\end{eqnarray}
The interval
\[
  \Delta \tau=p^6-p^{3/2}
\]
gives an estimation of the transition duration between asymptotic
waves.

We have developed the special asymptotic method which allows us to
receive the asymptotic solution of equations like
\[
  y''+\left(1+\frac{const}{x^{n}}+\frac{4}{x^{m}}\sin(2x+\al)\right)í=0
\]
with any desired order of accuracy. Formal scheme for construction
of general solution is similar with one for Mathieu-like equation.
At first we don't take into account the oscillating correction and
solve the oscillator-like equation. Then we add the mixed
harmonics into the solution, all constants before trigonometrical
functions replace by functions in the form of power series that
depends on $n$ and $m$, so the whole problem reduces to search of
expansion coefficients.

Let us apply this scheme to equation (\ref{Matie1}). We solve the
oscillator-like equation
\[
  \chi''+\left(1+\frac{p^2}{\tau^{4/3}}\right)\chi=0.
\]
Here $p^2/\tau^{4/3}\rightarrow 0$ steadily in the limit
$\tau\rightarrow \infty$, therefore the general asymptotic method
is applicable for solution of this equation (see Section 2). So we
have
\begin{eqnarray}
  \chi(\tau)=\frac1{\sqrt{1+\frac{p^2}{2\tau^{4/3}}}}\left[C_{1}
  e^{-i\left(\tau-\frac{3p^2}{2\tau^{1/3}}\right)}+C_{2}
  e^{i\left(\tau-\frac{3p^2}{2\tau^{1/3}}\right)}\right].
 \label{pervetap}
\end{eqnarray}
Now let's take into account the
influence of correction $(4/\tau) \sin(2\tau+\al)$. Equation
(\ref{Matie1}) one can write in a form
\begin{eqnarray}
  \chi''+\left(1+\frac{p^2}{\tau^{4/3}}+\frac{2i}{\tau}e^{i(2\tau+\al)}-
  \frac{2i}{\tau}e^{-i(2\tau+\al)}\right)\chi=0.
 \label{Matie2}
\end{eqnarray}
According to described scheme the general form of solution in the
limit $\tau\rightarrow \infty$ is
\begin{eqnarray}
\begin{array}{l} \di
  \chi(\tau)=\frac1{\sqrt{1+\frac{p^2}{2\tau^{4/3}}}}\sum_{s=0}^{\infty}
  \biggl[A_{s}(\tau)e^{i(2s+1)(\tau+\al/2)}e^{i\left(\frac{3p^2}{2\tau^{1/3}}+\vp_{1}\right)}+
  B_{s}(\tau)e^{-i(2s+1)(\tau+\al/2)}e^{-i\left(\frac{3p^2}{2\tau^{1/3}}+\vp_{2}\right)}+
\\[5mm]\di
  +C_{s}(\tau)e^{i(2s+1)(\tau+\al/2)}e^{-i\left(\frac{3p^2}{2\tau^{1/3}}+\vp_{2}\right)}+
  D_{s}(\tau)e^{-i(2s+1)(\tau+\al/2)}e^{i\left(\frac{3p^2}{2\tau^{1/3}}+\vp_{1}\right)}\biggr],
\end{array}
 \label{orlong}
\end{eqnarray} where $\vp_{1}, \; \vp_{2}$ = const.
In the frequency of equation (\ref{Matie2}) there are the terms
with powers of $\tau$ multiple to $1/3$. Therefore
\begin{eqnarray}
\begin{array}{l} \di
  A_{s}(\tau)=\sum_{k=3}^{-\infty}a_{sk}\tau^{k/3}, \quad
  B_{s}(\tau)=\sum_{k=3}^{-\infty}b_{sk}\tau^{k/3},
\\[5mm]\di
  C_{s}(\tau)=\sum_{k=3}^{-\infty}c_{sk}\tau^{k/3}, \quad
  D_{s}(\tau)=\sum_{k=3}^{-\infty}d_{sk}\tau^{k/3}.
\end{array}
 \label{coeflong}
\end{eqnarray} For a function
$\Psi_\vk(t)=\chi/(t-t_{0})$ we have obtained the following
expression:
\begin{eqnarray}
\begin{array}{l} \di
  \Psi_\vk(t)=\frac{C_{1}}{\sqrt{1+\frac{p^2}{2m^{4/3}(t-t_{0})^{4/3}}}}\biggl[\biggl(\left(
  \frac1{t-t_{0}}-\frac{20m^{2/3}}{3p^4(t-t_{0})^{1/3}}\right)
  \sin\left(\frac{3p^2}{2m^{1/3}(t-t_{0})^{1/3}}\right)-
\\[5mm] \di
  -\left(\frac{40m}{9p^6}-\frac{4m^{1/3}}{p^2(t-t_{0})^{2/3}}\right)
  \cos\left(\frac{3p^2}{2m^{1/3}(t-t_{0})^{1/3}}\right)\biggr)\sin(m(t-t_{0})+\al/2)+
\\[5mm] \di
  +\left(-\frac{2m^{1/3}}{p^2(t-t_{0})^{2/3}}\sin\left(\frac{3p^2}{2m^{1/3}(t-t_{0})^{1/3}}\right)+
  \left(\frac1{t-t_{0}}-\frac{4m^{2/3}}{3p^4(t-t_{0})^{1/3}}\right)
  \cos\left(\frac{3p^2}{2m^{1/3}(t-t_{0})^{1/3}}
  \right)\right)\times
\\[5mm] \di
  \times \cos(m(t-t_{0})+\al/2)\biggr]+
\end{array}
 \label{oresheniePsi}
\end{eqnarray}
\[
\begin{array}{l} \di
  \qquad\quad\;+\frac{C_{2}}{\sqrt{1+\frac{p^2}{2m^{4/3}(t-t_{0})^{4/3}}}}\biggl[\biggl(\left(
  \frac{20m^{2/3}}{3p^4(t-t_{0})^{1/3}}-\frac1{t-t_{0}}\right)
  \cos\left(\frac{3p^2}{2m^{1/3}(t-t_{0})^{1/3}}\right)-
\\[5mm] \di
  -\left(\frac{40m}{9p^6}-\frac{4m^{1/3}}{p^2(t-t_{0})^{2/3}}\right)
  \sin\left(\frac{3p^2}{2m^{1/3}(t-t_{0})^{1/3}}\right)\biggr)\sin(m(t-t_{0})+\al/2)+
\\[5mm] \di
  +\left(\frac{2m^{1/3}}{p^2(t-t_{0})^{2/3}}\cos\left(\frac{3p^2}{2m^{1/3}(t-t_{0})^{1/3}}\right)+
  \left(\frac1{t-t_{0}}-\frac{4m^{2/3}}{3p^4(t-t_{0})^{1/3}}\right)
  \sin\left(\frac{3p^2}{2m^{1/3}(t-t_{0})^{1/3}}
  \right)\right)\times
\\[5mm] \di
  \times \cos(m(t-t_{0})+\al/2)\biggr],
\end{array}
\]
where $\al$ is defined in (\ref{par}). Constants $C_1$ and $C_2$
are found by matching of solution (\ref{oresheniePsi}) with the
short-wave solution (\ref{shortresh}). The matching time is
defined from relation (\ref{csh}):
\begin{eqnarray}
  \tilde t(k)=\frac{p^{3/2}}{m}+t_{0}.
\end{eqnarray}
Therefore
\begin{eqnarray}
\begin{array}{l} \di
  \qquad\quad\;C_{1}=-\sqrt{\frac{3\varkappa}2}\frac{3H_{0}^2}
  {exp\left(\frac{9H_{0}^2}{4m^2}\right)m^{7/2}{\tilde{a}_{0}}^{3/2}}
  \biggl[\cos\left(-\frac{\al}{2}-\frac{107}{40}p^{3/2}-\frac1{2^{2/3}H_{0}}pm
  e^{\frac{3H_{0}^2}{2m^2}}\right)-
\\[8mm] \di
  \sin\left(-\frac{\al}{2}-\frac{107}{40}p^{3/2}-\frac1{2^{2/3}H_{0}}pm
  e^{\frac{3H_{0}^2}{2m^2}}\right)+
  7\cos\left(\frac{\al}{2}-\frac{147}{40}p^{3/2}-\frac1{2^{2/3}H_{0}}pm
  e^{\frac{3H_{0}^2}{2m^2}}\right)
\\[8mm] \di
  -7\sin\left(\frac{\al}{2}-\frac{147}{40}p^{3/2}-\frac1{2^{2/3}H_{0}}pm
  e^{\frac{3H_{0}^2}{2m^2}}\right)\biggr]\; ;
\end{array}
 \label{C1C2}
\end{eqnarray}
\[
\begin{array}{l} \di
  \qquad\quad\;C_{2}=-\sqrt{\frac{3\varkappa}2}\frac{3H_{0}^2}
  {exp\left(\frac{9H_{0}^2}{4m^2}\right)m^{7/2}{\tilde{a}_{0}}^{3/2}}
  \biggl[-7\sin\left(\frac{\al}{2}-\frac{147}{40}p^{3/2}-\frac1{2^{2/3}H_{0}}pm
  e^{\frac{3H_{0}^2}{2m^2}}\right)-
\\[8mm] \di
  7\cos\left(\frac{\al}{2}-\frac{147}{40}p^{3/2}-\frac1{2^{2/3}H_{0}}pm
  e^{\frac{3H_{0}^2}{2m^2}}\right)+
  \sin\left(-\frac{\al}{2}-\frac{107}{40}p^{3/2}-\frac1{2^{2/3}H_{0}}pm
  e^{\frac{3H_{0}^2}{2m^2}}\right)
\\[8mm] \di
  +\cos\left(-\frac{\al}{2}-\frac{107}{40}p^{3/2}-\frac1{2^{2/3}H_{0}}pm
  e^{\frac{3H_{0}^2}{2m^2}}\right)\biggr].
\end{array}
\]

Function of metric perturbations (\ref{oresheniePsi}) for every
regime can be essentially simplified:

1) Transient regime at $\tilde t(k)\lesssim t < t_{long},$ where
$t_{long}={p^6}/{m}+t_{0}$:
\begin{eqnarray}
  \Psi_\vk\approx\frac{C_{1}\cos\left(\frac{3p^2}{2m^{1/3}(t-t_{0})^{1/3}}-m(t-t_{0})-
  \frac{\al}{2}\right)+
  C_{2}\sin\left(\frac{3p^2}{2m^{1/3}(t-t_{0})^{1/3}}-m(t-t_{0})-\frac{\al}{2}\right)}{(t-t_{0})
  \sqrt{1+\frac{p^2}{2m^{4/3}(t-t_{0})^{4/3}}}}
 \label{perex}
\end{eqnarray}

2) Long-wave regime at $t > t_{long}$:
\begin{eqnarray}
  \Psi_\vk&\approx&-\frac{40m}{9p^6}\sin\left(m(t-t_{0})+\frac{\al}{2}\right)\times
 \nonumber
  \\
  &\times&
  \left[C_{1}\cos\left(\frac{3p^2}{2m^{1/3}(t-t_{0})^{1/3}}
  \right)+C_{2}\sin\left(\frac{3p^2}{2m^{1/3}(t-t_{0})^{1/3}}
  \right)\right]
 \label{longvoln}
\end{eqnarray}
Last expression one can also write in the form analogous to
(\ref{refpinf}):
\begin{eqnarray}
  \Psi_\vk\approx-\frac{40m}{9\sqrt{3}p^6}\frac{\sqrt{-\dot{H}}}{H}
  \left(C_{1}\cos\left(\frac{3p^2}{2m^{1/3}(t-t_{0})^{1/3}}
  \right)+C_{2}\sin\left(\frac{3p^2}{2m^{1/3}(t-t_{0})^{1/3}}
  \right)\right).
 \label{longvoln1}
\end{eqnarray}

Obviously the function of long-wave solution (\ref{longvoln})
oscillates with the constant amplitude as
$\sin\left(m(t-t_{0})+{\al}/{2}\right)$, however, in contrast to
(\ref{refpinf}), there is a slowly varying modulation factor,
which approaches to the constant just in the limit $t \rightarrow
\infty$.

Theoretical predictions for the parameter of energy density
perturbations spectrum are of the most interest. This parameter is
defined as follows
\begin{eqnarray}
 \Delta=\left( \bigl\langle 0| \left( \frac{\de\ve_{\vk\, {\rm inv}}}{\ve}
 \right)^2 | 0 \bigr\rangle \frac{k^3}{2\pi^2} \right)^{1/2}.
 \label{HZs}
\end{eqnarray}
The analytical results of calculation of $\Delta$ without the
averaging over the initial phases of quantum fluctuations are
shown in Appendix.

\section{Conclusion.}

Gauge invariant approach to investigation of linear scalar
perturbations of inflaton and gravitational fields (SIG-waves) has
been developed. We mean the derivation and analytical
investigation of equation for the single invariant function
$J_{\vk}$ without resorting any gauge. Equation of invariant
dynamics (EID) is constructed both in the cosmic and conformal
times. The transformations of EID coefficients together with
background equations for arbitrary potential have shown an absence
of the mathematical singularities in reality. This fact has given
a chance for analytical investigations of the EID. This internal
property of the invariant dynamics is its significant advantage.

Our approach allows to compare various gauges used by other
researchers, and to find unambiguous selection criteria of
physical and coordinate effects. We have shown that the so-called
longitudinal gauge commonly used for studying the gravitational
instability leads to the overestimation of physical effects due to
presence of nonphysical proper time perturbations in results
obtained by using this gauge.

We have proven that the non-invariant part of energy density
perturbations in the synchronous gauge is equal to zero for any
co-ordinates transformations preserving the gauge, therefore the
calculation of the relative perturbation of energy density
$\de\ve/\ve$ in the synchronous gauge gives correct results in
contrast to the longitudinal one.

To a great extent, the technology of EID solving rests upon
mathematical symmetry properties of invariant dynamics. The
general long-wave solution of EID as a functional of background
solution for arbitrary potential $U(\phi)$ is obtained. We have
developed asymptotical methods which allow us to obtain the
invariant function of metric perturbations in analytical form at
all stages of the Universe evolution for arbitrary wave lengths in
the framework of fixed potential (we use of the simplest model
with $U(\vp)=m^{2}\vp^{2}/2$). It is important point of our
investigation that both the non-perturbed or background
characteristics (Hubble function $H(t),$ inflaton $\phi(t)$ and
scalar factor $a(t)$) and characteristics of metric perturbations
(invariant functions $J_\vk$ or $\Psi_\vk$) at the inflationary
and postinflationary stages in the considering model
(\ref{lagran}) of the Universe are completely defined by parameter
$m$ and initial values of Hubble function $H_{0}$ and scale factor
$\tilde{a}_0.$ Note also that all analytical results coincide with
numerical solution of EID at various stages and wave numbers with
a high precision.

We have also obtained analytical expressions for the energy
density perturbations spectrum $\Delta(k,t)$ for all possible wave
numbers $k$ and times $t.$ (see Appendix). Amplitude of the
long-wave spectrum in the case of transition from short waves to
long ones occurs at the inflationary stage is almost flat, i.e.
has the Harrison-Zeldovich form, for arbitrary potential $U(\vp),$
but there is some tilt that can be compared with one from recent
high precision data. This result is supposed to be the principal
result of our invariant approach. Inflationary prediction for
nearly flat spectrum of density perturbations is in agreement with
both measurements of the CMB anisotropy and observations of
structures in the Universe.

We are grateful to V.A. Beylin, V.I. Kuksa, O.D. Lalakulich, L.S.
Marochnik, A.V. Zayakin and O.V. Teryaev for useful discussions
and comments. This work is partially supported by grant RFBR
03-02-16816.

\section{Appendix. Calculation of the energy density perturbation spectrum.}

In this section there are the analytical expressions for the
parameter of energy density perturbations spectrum $\Delta(k,t)$
(\ref{HZs}):
\begin{eqnarray}
  \Delta(k,t)=\frac{k^{3/2}}{\sqrt{2}\pi}\left\vert
  \frac{\de\ve_{inv}}{\ve}\right\vert.
\end{eqnarray}
Expression for
relative perturbations of energy density through $\Psi_\vk$ has a
form:
\begin{eqnarray}
  \frac{\de\ve_{inv}}{\ve}=-\frac{\sqrt{-2\dot{H}}}{3H^2}\left[
  \dot{\Psi}_\vk+\left(\frac{\dot H}{H}-3H-\frac{\ddot H}{2\dot
  H}\right)\Psi_\vk \right] +\frac{\sqrt{2}\dot{H}}{H}
  \int\limits_{t_0}^{t} \frac{\sqrt{-\dot{H}}}{H} \Psi_\vk dt.
\end{eqnarray}
Using the expressions for $\Psi_\vk$ at various $t$ and $k$
(\ref{refinf}), (\ref{refpinf}), (\ref{shortresh}), (\ref{perex})
and (\ref{longvoln1}), we have obtained the following expressions
for ${\de\ve_{inv}}/{\ve}$ and $\Delta(k,t)$:

\begin{enumerate}

\item {\it Short waves at the inflationary stage}
that is at times $0<t<\tilde{t}(k),\quad
t\leqslant t_{1},$ where $\tilde{t}(k)$ is the transition time
from short waves to long ones, $t_{1}$ is the time when the
inflationary stage terminates. In that case we have
\begin{eqnarray}
\begin{array}{l} \di
  \frac{\de\ve_{inv}}{\ve}=-\frac{2\sqrt{3\varkappa}mH_{0}^2}{9{H(t)}^3{\tilde{a}_{0}}^2(H_{0}+H(t))}
  k^{1/2}e^{-t(H_{0}+H(t))}\times
\\[5mm] \di
  \times\left[\cos\left(\frac{2ke^{-\frac{t(H_{0}+H(t))}{2}}}{(H_{0}+H(t))\tilde{a}_{0}}
  -\frac{k}{H_{0}\tilde{a}_{0}}\right)+\sin\left(\frac{2ke^{-\frac{t(H_{0}+H(t))}{2}}}{(H_{0}+H(t))\tilde{a}_{0}}-
  \frac{k}{H_{0}\tilde{a}_{0}}\right)\right];
\end{array}
 \label{dee1}
\end{eqnarray}
\begin{eqnarray}
\begin{array}{l} \di
  \Delta(k,t)=\frac{\sqrt{6\varkappa}mH_{0}^2}{9\pi{H(t)}^3{\tilde{a}_{0}}^2(H_{0}+H(t))}
  k^{2}e^{-t(H_{0}+H(t))}\times
\\[5mm] \di
  \times\left\vert
  \cos\left(\frac{2ke^{-\frac{t(H_{0}+H(t))}{2}}}{(H_{0}+H(t))\tilde{a}_{0}}-
  \frac{k}{H_{0}\tilde{a}_{0}}\right)+\sin\left(\frac{2ke^{-\frac{t(H_{0}+H(t))}{2}}}{(H_{0}+H(t))\tilde{a}_{0}}-
  \frac{k}{H_{0}\tilde{a}_{0}}\right)\right\vert.
\end{array}
 \label{delta1}
\end{eqnarray}

\item {\it Long waves in the case of the transition from short waves to long ones occurs at the
inflationary stage} that is at times $t>\tilde{t}(k),\quad
\tilde{t}(k)\leqslant t_{1}$. Here we have
\begin{eqnarray}
  \frac{\de\ve_{inv}}{\ve}=\frac{\sqrt{3\varkappa}H_{0}^2}{mH_{s}}\frac{\dot{H}(t)}{k^{3/2}H(t)}
  \left[\cos\left(\frac{k}{H_{0}\tilde{a}_{0}}\right)-\sin\left(\frac{k}{H_{0}\tilde{a}_{0}}\right)\right];
\end{eqnarray}
\begin{eqnarray}
  \Delta(k,t)=-\sqrt{\frac{3\varkappa}2}\frac{H_{0}^2}{\pi
  mH_{s}}\frac{\dot{H}(t)}{H(t)} \left\vert
  \cos\left(\frac{k}{H_{0}\tilde{a}_{0}}\right)-\sin\left(\frac{k}{H_{0}\tilde{a}_{0}}\right)\right\vert,
\end{eqnarray} where
\[
  H_{s}\equiv H(\tilde
  t(k))=\sqrt{{H_{0}}^2-\frac{2m^2}{3}\;\ln\left(\frac{k}{m\tilde{a}_{0}}\right)}
\]
is the value of Hubble function at the transition moment. We see
that the amplitude of this spectrum is almost constant with $k$
for arbitrary potential $U(\vp).$ Therefore this spectrum has the
Harrison-Zeldovich form. Some tilt from flatness contains in $H_s$
and can be compared with recent high precision data.

\item {\it Long waves in the case of the transition from short waves to long ones occurs at the
postinflationary stage} that is at times $t>t_{long},\quad
t_{long}>t_{1}:$
\begin{eqnarray}
\begin{array}{l} \di
 \qquad\frac{\de\ve_{inv}}{\ve}=-\frac{20\sqrt{\varkappa}H_{0}^2}
 {m^{7/2}{\tilde{a}_{0}}^{3/2}exp\left(\frac{9H_{0}^2}{4m^2}\right)}
 \frac{\dot{H}(t)}{H(t)}
 \biggl[\cos\left(\frac{3p^2}{2m^{1/3}(t-t_{0})^{1/3}}
 \right)\times
\\[5mm] \di
 \times\biggl(\cos\left(-\frac{\al}{2}-\frac{107}{40}p^{3/2}-\frac1{2^{2/3}H_{0}}pm
 e^{\frac{3H_{0}^2}{2m^2}}\right)-
 \sin\left(-\frac{\al}{2}-\frac{107}{40}p^{3/2}-\frac1{2^{2/3}H_{0}}pm
 e^{\frac{3H_{0}^2}{2m^2}}\right)+
\\[5mm] \di
 7\cos\left(\frac{\al}{2}-\frac{147}{40}p^{3/2}-\frac1{2^{2/3}H_{0}}pm
 e^{\frac{3H_{0}^2}{2m^2}}\right)
 -7\sin\left(\frac{\al}{2}-\frac{147}{40}p^{3/2}-\frac1{2^{2/3}H_{0}}pm
 e^{\frac{3H_{0}^2}{2m^2}}\right)\biggr)+
\end{array}
\end{eqnarray}
\[
\begin{array}{l} \di
 \qquad+\sin\left(\frac{3p^2}{2m^{1/3}(t-t_{0})^{1/3}}
 \right)\biggl(-7\sin\left(\frac{\al}{2}-\frac{147}{40}p^{3/2}-\frac1{2^{2/3}H_{0}}pm
 e^{\frac{3H_{0}^2}{2m^2}}\right)-
\\[5mm] \di
 7\cos\left(\frac{\al}{2}-\frac{147}{40}p^{3/2}-\frac1{2^{2/3}H_{0}}pm
 e^{\frac{3H_{0}^2}{2m^2}}\right)+
 \sin\left(-\frac{\al}{2}-\frac{107}{40}p^{3/2}-\frac1{2^{2/3}H_{0}}pm
 e^{\frac{3H_{0}^2}{2m^2}}\right)
\\[5mm] \di
 +\cos\left(-\frac{\al}{2}-\frac{107}{40}p^{3/2}-\frac1{2^{2/3}H_{0}}pm
 e^{\frac{3H_{0}^2}{2m^2}}\right)\biggr)\biggr];
\end{array}
\]

\begin{eqnarray}
\begin{array}{l} \di
 \qquad\Delta(p,t)=-\frac{5\sqrt{2\varkappa}H_{0}^2} {\pi
 m^{2}}p^{3/2} \frac{\dot{H}(t)}{H(t)} \biggl\vert
 \cos\left(\frac{3p^2}{2m^{1/3}(t-t_{0})^{1/3}} \right)\times
\\[5mm] \di
 \times\biggl(\cos\left(-\frac{\al}{2}-\frac{107}{40}p^{3/2}-\frac1{2^{2/3}H_{0}}pm
 e^{\frac{3H_{0}^2}{2m^2}}\right)-
 \sin\left(-\frac{\al}{2}-\frac{107}{40}p^{3/2}-\frac1{2^{2/3}H_{0}}pm
 e^{\frac{3H_{0}^2}{2m^2}}\right)+
\\[5mm] \di
 7\cos\left(\frac{\al}{2}-\frac{147}{40}p^{3/2}-\frac1{2^{2/3}H_{0}}pm
 e^{\frac{3H_{0}^2}{2m^2}}\right)
 -7\sin\left(\frac{\al}{2}-\frac{147}{40}p^{3/2}-\frac1{2^{2/3}H_{0}}pm
 e^{\frac{3H_{0}^2}{2m^2}}\right)\biggr)+
\end{array}
\end{eqnarray}
\[
\begin{array}{l} \di
 \qquad+\sin\left(\frac{3p^2}{2m^{1/3}(t-t_{0})^{1/3}}
 \right)\biggl(-7\sin\left(\frac{\al}{2}-\frac{147}{40}p^{3/2}-\frac1{2^{2/3}H_{0}}pm
 e^{\frac{3H_{0}^2}{2m^2}}\right)-
\\[5mm] \di
 7\cos\left(\frac{\al}{2}-\frac{147}{40}p^{3/2}-\frac1{2^{2/3}H_{0}}pm
 e^{\frac{3H_{0}^2}{2m^2}}\right)+
 \sin\left(-\frac{\al}{2}-\frac{107}{40}p^{3/2}-\frac1{2^{2/3}H_{0}}pm
 e^{\frac{3H_{0}^2}{2m^2}}\right)
\\[5mm] \di
 +\cos\left(-\frac{\al}{2}-\frac{107}{40}p^{3/2}-\frac1{2^{2/3}H_{0}}pm
 e^{\frac{3H_{0}^2}{2m^2}}\right)\biggr)\biggr\vert.
 \end{array}
\]
Here instead of $k$ we introduced a new parameter
\[
  p=p(k)=\frac{2^{2/3}k}{\tilde{a}_{0}mexp\left(\frac{3H_{0}^2}{2m^2}\right)}\gg
  2^{2/3}.
\]
Since these perturbations are considered at the postinflationary
stage so in the framework of investigated model
\[
  \frac{\dot{H}(t)}{H(t)}\approx
  -\frac2{t-t_{0}}\sin^2\left(m(t-t_{0})+ \frac{\al}{2}\right).
\]
Apparently, the long-wave functions $\de\ve_{inv}/\ve$ and
$\Delta(k,t)$ are functionals of background solution $H(t)$ and
depend on time as ${\dot{H}(t)}/{H(t)}$. This statement is correct
for any potential.

\item {\it Short waves at the postinflationary stage}
that is at times $t_{1}<t<\tilde t(k), \quad p\gg 2^{2/3}$:
\begin{eqnarray}
\begin{array}{l} \di
\frac{\de\ve_{inv}}{\ve}=-\sqrt{3\varkappa}
\frac{9H_{0}^2}{m^{19/6}{\tilde{a}_{0}}^{3/2}
exp\left(\frac{9H_{0}^2}{4m^2}\right)\sqrt{1+\frac{m^{4/3}}{2p^2}(t-t_{0})^{4/3}}}
\times
\\[5mm] \di
\times\biggl[\frac{p^{1/2}m^{1/3}}{(t-t_{0})^{1/3}}\biggl(\left(\cos\left(\frac{pm}{2^{2/3}H_{0}}
e^{\frac{3H_{0}^2}{2m^2}}\right)+\sin\left(\frac{pm}{2^{2/3}H_{0}}
e^{\frac{3H_{0}^2}{2m^2}}\right)\right)\times
\\[5mm] \di
\times\left(\cos\left(m(t-t_{0})-3pm^{1/3}(t-t_{0})^{1/3}+\frac{\al}{2}\right)-
\cos\left(m(t-t_{0})+3pm^{1/3}(t-t_{0})^{1/3}+\frac{\al}{2}\right)\right)-
\\[5mm] \di
\qquad\qquad\quad-\left(\cos\left(\frac{pm}{2^{2/3}H_{0}}
e^{\frac{3H_{0}^2}{2m^2}}\right)-\sin\left(\frac{pm}{2^{2/3}H_{0}}
e^{\frac{3H_{0}^2}{2m^2}}\right)\right)\times
\\[5mm] \di
\times\left(\sin\left(m(t-t_{0})+3pm^{1/3}(t-t_{0})^{1/3}+\frac{\al}{2}\right)+
\sin\left(m(t-t_{0})-3pm^{1/3}(t-t_{0})^{1/3}+\frac{\al}{2}\right)\right)\biggr)+
\end{array}
\end{eqnarray}
\[
\begin{array}{l} \di
+\frac{m(t-t_{0})^{1/3}}{p^{1/2}}\biggl(\left(\cos\left(\frac{pm}{2^{2/3}H_{0}}
e^{\frac{3H_{0}^2}{2m^2}}\right)+\sin\left(\frac{pm}{2^{2/3}H_{0}}
e^{\frac{3H_{0}^2}{2m^2}}\right)\right)\times
\\[5mm] \di
\times\left(\cos\left(m(t-t_{0})-3pm^{1/3}(t-t_{0})^{1/3}+\frac{\al}{2}\right)+
\cos\left(m(t-t_{0})+3pm^{1/3}(t-t_{0})^{1/3}+\frac{\al}{2}\right)\right)+
\\[5mm] \di
\qquad\qquad\qquad+\left(\cos\left(\frac{pm}{2^{2/3}H_{0}}
e^{\frac{3H_{0}^2}{2m^2}}\right)-\sin\left(\frac{pm}{2^{2/3}H_{0}}
e^{\frac{3H_{0}^2}{2m^2}}\right)\right)\times
\\[5mm] \di
\times\left(\sin\left(m(t-t_{0})+3pm^{1/3}(t-t_{0})^{1/3}+\frac{\al}{2}\right)-
\sin\left(m(t-t_{0})-3pm^{1/3}(t-t_{0})^{1/3}+\frac{\al}{2}\right)\right)\biggr)\biggr];
\end{array}
\]

\begin{eqnarray}
\begin{array}{l} \di
\Delta(p,t)=\sqrt{\frac{3\varkappa}2} \frac{9H_{0}^2}{2\pi
m^{5/3}\sqrt{1+\frac{m^{4/3}}{2p^2}(t-t_{0})^{4/3}}} \times
\\[5mm] \di
\times\biggl\vert\frac{p^{2}m^{1/3}}{(t-t_{0})^{1/3}}\biggl(\left(\cos\left(\frac{pm}{2^{2/3}H_{0}}
e^{\frac{3H_{0}^2}{2m^2}}\right)+\sin\left(\frac{pm}{2^{2/3}H_{0}}
e^{\frac{3H_{0}^2}{2m^2}}\right)\right)\times
\\[5mm] \di
\times\left(\cos\left(m(t-t_{0})-3pm^{1/3}(t-t_{0})^{1/3}+\frac{\al}{2}\right)-
\cos\left(m(t-t_{0})+3pm^{1/3}(t-t_{0})^{1/3}+\frac{\al}{2}\right)\right)-
\\[5mm] \di
\qquad\qquad\quad-\left(\cos\left(\frac{pm}{2^{2/3}H_{0}}
e^{\frac{3H_{0}^2}{2m^2}}\right)-\sin\left(\frac{pm}{2^{2/3}H_{0}}
e^{\frac{3H_{0}^2}{2m^2}}\right)\right)\times
\\[5mm] \di
\times\left(\sin\left(m(t-t_{0})+3pm^{1/3}(t-t_{0})^{1/3}+\frac{\al}{2}\right)+
\sin\left(m(t-t_{0})-3pm^{1/3}(t-t_{0})^{1/3}+\frac{\al}{2}\right)\right)\biggr)+
\end{array}
\end{eqnarray}
\[
\begin{array}{l} \di
+pm(t-t_{0})^{1/3}\biggl(\left(\cos\left(\frac{pm}{2^{2/3}H_{0}}
e^{\frac{3H_{0}^2}{2m^2}}\right)+\sin\left(\frac{pm}{2^{2/3}H_{0}}
e^{\frac{3H_{0}^2}{2m^2}}\right)\right)\times
\\[5mm] \di
\times\left(\cos\left(m(t-t_{0})-3pm^{1/3}(t-t_{0})^{1/3}+\frac{\al}{2}\right)+
\cos\left(m(t-t_{0})+3pm^{1/3}(t-t_{0})^{1/3}+\frac{\al}{2}\right)\right)+
\\[5mm] \di
\qquad\qquad\qquad+\left(\cos\left(\frac{pm}{2^{2/3}H_{0}}
e^{\frac{3H_{0}^2}{2m^2}}\right)-\sin\left(\frac{pm}{2^{2/3}H_{0}}
e^{\frac{3H_{0}^2}{2m^2}}\right)\right)\times
\\[5mm] \di
\times\left(\sin\left(m(t-t_{0})+3pm^{1/3}(t-t_{0})^{1/3}+\frac{\al}{2}\right)-
\sin\left(m(t-t_{0})-3pm^{1/3}(t-t_{0})^{1/3}+\frac{\al}{2}\right)\right)\biggr)\biggr\vert.
\end{array}
\]

\item {\it Waves of middle lengths} (occurs only at the postinflationary stage)
that is at times $t_{1}<\tilde t(k)<t<t_{long}, \quad p\gg
2^{2/3}$:
\begin{eqnarray}
\begin{array}{l} \di
\frac{\de\ve_{inv}}{\ve}=-\sqrt{\varkappa}\frac{9H_{0}^2}
{2m^{7/2}{\tilde{a}_{0}}^{3/2}exp\left(\frac{9H_{0}^2}{4m^2}\right)
\sqrt{1+\frac{p^2}{2m^{4/3}(t-t_{0})^{4/3}}}}\times
\\[8mm] \di
\times\biggl[\frac{p^2}{4m^{1/3}(t-t_{0})^{4/3}}
\biggl(\biggl[\cos\left(-\frac{\al}{2}-\frac{107}{40}p^{3/2}-\frac1{2^{2/3}H_{0}}pm
e^{\frac{3H_{0}^2}{2m^2}}\right)-
\\[8mm] \di
-\sin\left(-\frac{\al}{2}-\frac{107}{40}p^{3/2}-\frac1{2^{2/3}H_{0}}pm
e^{\frac{3H_{0}^2}{2m^2}}\right)+
7\cos\left(\frac{\al}{2}-\frac{147}{40}p^{3/2}-\frac1{2^{2/3}H_{0}}pm
e^{\frac{3H_{0}^2}{2m^2}}\right)-
\\[8mm] \di
\qquad\qquad\quad-7\sin\left(\frac{\al}{2}-\frac{147}{40}p^{3/2}-\frac1{2^{2/3}H_{0}}pm
e^{\frac{3H_{0}^2}{2m^2}}\right)\biggr]\times
\\[8mm] \di
\times\left(\cos\left(\frac{3p^2}{2m^{1/3}(t-t_{0})^{1/3}}\right)-
\cos\left(2m(t-t_{0})-\frac{3p^2}{2m^{1/3}(t-t_{0})^{1/3}}+\al\right)\right)+
\\[8mm] \di
\qquad\qquad\quad+\biggl[-7\sin\left(\frac{\al}{2}-\frac{147}{40}p^{3/2}-\frac1{2^{2/3}H_{0}}pm
e^{\frac{3H_{0}^2}{2m^2}}\right)-
\\[8mm] \di
-7\cos\left(\frac{\al}{2}-\frac{147}{40}p^{3/2}-\frac1{2^{2/3}H_{0}}pm
e^{\frac{3H_{0}^2}{2m^2}}\right)+
\sin\left(-\frac{\al}{2}-\frac{107}{40}p^{3/2}-\frac1{2^{2/3}H_{0}}pm
e^{\frac{3H_{0}^2}{2m^2}}\right)+
\\[8mm] \di
\qquad\qquad\quad+\cos\left(-\frac{\al}{2}-\frac{107}{40}p^{3/2}-\frac1{2^{2/3}H_{0}}pm
e^{\frac{3H_{0}^2}{2m^2}}\right)\biggr]\times
\\[8mm] \di
\times\left(\sin\left(\frac{3p^2}{2m^{1/3}(t-t_{0})^{1/3}}\right)+
\sin\left(2m(t-t_{0})-\frac{3p^2}{2m^{1/3}(t-t_{0})^{1/3}}+\al\right)\right)\biggr)+
\\[8mm] \di
+m\biggl(\biggl[\cos\left(-\frac{\al}{2}-\frac{107}{40}p^{3/2}-\frac1{2^{2/3}H_{0}}pm
e^{\frac{3H_{0}^2}{2m^2}}\right)-
\\[8mm] \di
-\sin\left(-\frac{\al}{2}-\frac{107}{40}p^{3/2}-\frac1{2^{2/3}H_{0}}pm
e^{\frac{3H_{0}^2}{2m^2}}\right)+
7\cos\left(\frac{\al}{2}-\frac{147}{40}p^{3/2}-\frac1{2^{2/3}H_{0}}pm
e^{\frac{3H_{0}^2}{2m^2}}\right)-
\\[8mm] \di
-7\sin\left(\frac{\al}{2}-\frac{147}{40}p^{3/2}-\frac1{2^{2/3}H_{0}}pm
e^{\frac{3H_{0}^2}{2m^2}}\right)\biggr]
\cos\left(\frac{3p^2}{2m^{1/3}(t-t_{0})^{1/3}}\right)+
\\[8mm] \di
\qquad\qquad\quad+\biggl[-7\sin\left(\frac{\al}{2}-\frac{147}{40}p^{3/2}-\frac1{2^{2/3}H_{0}}pm
e^{\frac{3H_{0}^2}{2m^2}}\right)-
\\[8mm] \di
-7\cos\left(\frac{\al}{2}-\frac{147}{40}p^{3/2}-\frac1{2^{2/3}H_{0}}pm
e^{\frac{3H_{0}^2}{2m^2}}\right)+
\sin\left(-\frac{\al}{2}-\frac{107}{40}p^{3/2}-\frac1{2^{2/3}H_{0}}pm
e^{\frac{3H_{0}^2}{2m^2}}\right)+
\\[8mm] \di
+\cos\left(-\frac{\al}{2}-\frac{107}{40}p^{3/2}-\frac1{2^{2/3}H_{0}}pm
e^{\frac{3H_{0}^2}{2m^2}}\right)\biggr]
\sin\left(\frac{3p^2}{2m^{1/3}(t-t_{0})^{1/3}}\right)\biggr)\biggr];
\end{array}
\end{eqnarray}

\begin{eqnarray}
\begin{array}{l} \di
\Delta(p,t)=\sqrt{\frac{\varkappa}2}\frac{9H_{0}^2} {4\pi
m^{2}\sqrt{1+\frac{p^2}{2m^{4/3}(t-t_{0})^{4/3}}}}\times
\\[8mm] \di
\times\biggl\vert\frac{p^{7/2}}{4m^{1/3}(t-t_{0})^{4/3}}
\biggl(\biggl[\cos\left(-\frac{\al}{2}-\frac{107}{40}p^{3/2}-\frac1{2^{2/3}H_{0}}pm
e^{\frac{3H_{0}^2}{2m^2}}\right)-
\\[8mm] \di
-\sin\left(-\frac{\al}{2}-\frac{107}{40}p^{3/2}-\frac1{2^{2/3}H_{0}}pm
e^{\frac{3H_{0}^2}{2m^2}}\right)+
7\cos\left(\frac{\al}{2}-\frac{147}{40}p^{3/2}-\frac1{2^{2/3}H_{0}}pm
e^{\frac{3H_{0}^2}{2m^2}}\right)-
\\[8mm] \di
\qquad\qquad\quad-7\sin\left(\frac{\al}{2}-\frac{147}{40}p^{3/2}-\frac1{2^{2/3}H_{0}}pm
e^{\frac{3H_{0}^2}{2m^2}}\right)\biggr]\times
\\[8mm] \di
\times\left(\cos\left(\frac{3p^2}{2m^{1/3}(t-t_{0})^{1/3}}\right)-
\cos\left(2m(t-t_{0})-\frac{3p^2}{2m^{1/3}(t-t_{0})^{1/3}}+\al\right)\right)+
\\[8mm] \di
\qquad\qquad\quad+\biggl[-7\sin\left(\frac{\al}{2}-\frac{147}{40}p^{3/2}-\frac1{2^{2/3}H_{0}}pm
e^{\frac{3H_{0}^2}{2m^2}}\right)-
\\[8mm] \di
-7\cos\left(\frac{\al}{2}-\frac{147}{40}p^{3/2}-\frac1{2^{2/3}H_{0}}pm
e^{\frac{3H_{0}^2}{2m^2}}\right)+
\sin\left(-\frac{\al}{2}-\frac{107}{40}p^{3/2}-\frac1{2^{2/3}H_{0}}pm
e^{\frac{3H_{0}^2}{2m^2}}\right)+
\\[8mm] \di
\qquad\qquad\quad+\cos\left(-\frac{\al}{2}-\frac{107}{40}p^{3/2}-\frac1{2^{2/3}H_{0}}pm
e^{\frac{3H_{0}^2}{2m^2}}\right)\biggr]\times
\\[8mm] \di
\times\left(\sin\left(\frac{3p^2}{2m^{1/3}(t-t_{0})^{1/3}}\right)+
\sin\left(2m(t-t_{0})-\frac{3p^2}{2m^{1/3}(t-t_{0})^{1/3}}+\al\right)\right)\biggr)+
\\[8mm] \di
+p^{3/2}m\biggl(\biggl[\cos\left(-\frac{\al}{2}-\frac{107}{40}p^{3/2}-\frac1{2^{2/3}H_{0}}pm
e^{\frac{3H_{0}^2}{2m^2}}\right)-
\\[8mm] \di
-\sin\left(-\frac{\al}{2}-\frac{107}{40}p^{3/2}-\frac1{2^{2/3}H_{0}}pm
e^{\frac{3H_{0}^2}{2m^2}}\right)+
7\cos\left(\frac{\al}{2}-\frac{147}{40}p^{3/2}-\frac1{2^{2/3}H_{0}}pm
e^{\frac{3H_{0}^2}{2m^2}}\right)-
\\[8mm] \di
-7\sin\left(\frac{\al}{2}-\frac{147}{40}p^{3/2}-\frac1{2^{2/3}H_{0}}pm
e^{\frac{3H_{0}^2}{2m^2}}\right)\biggr]
\cos\left(\frac{3p^2}{2m^{1/3}(t-t_{0})^{1/3}}\right)+
\\[8mm] \di
\qquad\qquad\quad+\biggl[-7\sin\left(\frac{\al}{2}-\frac{147}{40}p^{3/2}-\frac1{2^{2/3}H_{0}}pm
e^{\frac{3H_{0}^2}{2m^2}}\right)-
\\[8mm] \di
-7\cos\left(\frac{\al}{2}-\frac{147}{40}p^{3/2}-\frac1{2^{2/3}H_{0}}pm
e^{\frac{3H_{0}^2}{2m^2}}\right)+
\sin\left(-\frac{\al}{2}-\frac{107}{40}p^{3/2}-\frac1{2^{2/3}H_{0}}pm
e^{\frac{3H_{0}^2}{2m^2}}\right)+
\\[8mm] \di
+\cos\left(-\frac{\al}{2}-\frac{107}{40}p^{3/2}-\frac1{2^{2/3}H_{0}}pm
e^{\frac{3H_{0}^2}{2m^2}}\right)\biggr]
\sin\left(\frac{3p^2}{2m^{1/3}(t-t_{0})^{1/3}}\right)\biggr)\biggr\vert.
\end{array}
\end{eqnarray}

\end{enumerate}

\end{document}